\documentclass[proceedings]{JHEP}

\usepackage{epsfig}                     

\newbox\mybox
\newcommand\fverb{\setbox\mybox=\hbox\bgroup\verb}
\newcommand\fverbdo{\egroup\medskip\noindent\fbox{\unhbox\mybox}\ }
\newcommand\fverbit{\egroup\item[\fbox{\unhbox\mybox}]}

\font\beeg=cmr17 scaled 1600            
\newcommand\init[1]{\setbox\mybox=\hbox{{\beeg #1}~}%
                   \noindent\global\hangindent=\wd\mybox\global\hangafter-2%
                   \sc\smash{\llap {\lower 13.2pt \box\mybox}}}

\title{Perturbative heavy quark-antiquark systems}

\author{M. Beneke\\ 
        Theory Division, CERN, CH-1211 Geneva 23, Switzerland\\
        E-mail: \email{martin.beneke@cern.ch}}

\conference{Heavy Flavours 8, Southampton, UK, 1999}

\abstract{In this review I cover recent developments concerning 
the construction of non-relativistic effective theories for perturbative 
heavy quark-antiquark systems and heavy quark mass definitions. I then 
discuss next-to-next-to-leading order results on quarkonium masses and 
decay, top quark pair production near threshold and QCD sum rules for 
$\Upsilon$ mesons.
\centerline{(November 1999, CERN-TH/99-355, hep-ph/9911490)}
}

\begin{document} 

\vspace*{-1cm}
{\init Heavy quark} effective field theory  
has profoundly influenced the way we think about mesons made of a heavy 
and a light quark \cite{hqet}. 
It provides a systematic expansion in powers of 
$\alpha_s$ and $\Lambda_{\rm QCD}/m_Q$. Perhaps more important, it separates 
hard (momenta $l\sim m_Q$) and soft ($l\sim \Lambda_{\rm QCD}$) 
physics. Hard effects can be calculated; spin-flavour symmetry \cite{IW89} 
relates soft effects and makes the framework predictive. 

The construction of an effective theory for mesons made of a heavy quark 
and antiquark (``onia'') turned out to be more difficult. There are 
important differences between $Q\bar{Q}$ systems and $Q\bar{q}$ systems: 
the former develop weak-coup\-ling bound states in the heavy quark limit, 
the latter don't. The expansion parameter for onia is the velocity of the 
heavy quark, $v$, rather than $\Lambda_{\rm QCD}/m_Q$. There is no 
flavour symmetry. But first of all, the presence of four momentum scales 
$m_Q$, $m_Q v$, $m_Q v^2$ and $\Lambda_{\rm QCD}$ rather than only two 
complicates the effective theory construction.

In a seminal paper Caswell and Lepage introduced effective field theory 
methods to QED bound state calculations \cite{CL86}. 
The QCD equivalent, called non-relativistic QCD (NRQCD) \cite{nrqcd}, 
has now become an established tool 
in lattice simulations and for describing quarkonium production and 
decay \cite{BBL95}. However, for perturbative heavy quark-antiquark 
systems, by which I mean onia that satisfy $m_Q v^2\gg \Lambda_{\rm QCD}$ 
in addition to $v\ll 1$, NRQCD is not yet optimal. 
NRQCD factorizes the scale $m_Q$ (and needs to assume only that $m_Q\gg 
\Lambda_{\rm QCD}$), but does not deal with the large scale hierarchy 
$m_Q v\gg m_Q v^2$. It does not make explicit the dominance of the static 
Coulomb force and the approximate 
quantum-mechanical nature of perturbative $Q\bar{Q}$ systems.

During the past two years progress on perturbative 
$Q\bar{Q}$ systems has been rapid. 
Some details remain to be clarified, but the effective field theory 
(EFT) picture is now essentially complete. Several advanced applications 
have been worked out. The use of effective theory may not be 
compulsory for perturbative $Q\bar{Q}$ systems. However, the 
gain in systematics in the expansion in $v$, 
transparency of language and, eventually, 
technical simplification, is enormous. In this sense effective field 
theory has had as profound an impact on understanding onia 
as it had on understanding heavy-light mesons.

\section{Effective theories}

\subsection{NRQCD}

Scattering processes at momentum transfer much smaller than $m_Q$ 
are reproduced by the non-relati\-vi\-stic Lagrangian \cite{nrqcd} 
\begin{eqnarray}
\label{nrqcd}
{\cal L}_{\rm NRQCD} &=& 
\psi^\dagger \left(i D^0+\frac{\vec{D}^2}{2 m_Q}
\right)\psi + \frac{1}{8 m_Q^3}\,\psi^\dagger\vec{D}^4\psi
\nonumber\\
&&\hspace*{-1.5cm}
-\,\frac{d_1\,g_s}{2 m_Q}\,\psi^\dagger\vec{\sigma}\cdot \vec{B}\psi
+\frac{d_2\,g_s}{8 m_Q^2}\,\psi^\dagger\left(\vec{D}\cdot\vec{E}-
\vec{E}\cdot\vec{D}\right)\psi
\nonumber\\
&&\hspace*{-1.5cm}
+ \,\frac{d_3\,i g_s}{8 m_Q^2}\,\psi^\dagger\vec{\sigma}\cdot\left(
\vec{D}\times\vec{E}-\vec{E}\times\vec{D}\right)\psi+\ldots  
\nonumber\\[0.1cm]
&&\hspace*{-1.5cm}
+ \,\mbox{antiquark terms}
\nonumber\\
&&\hspace*{-1.5cm}
\,+ \sum_\Gamma \frac{d_\Gamma}{m_Q^2}(\psi^\dagger\Gamma\chi)
(\chi^\dagger\Gamma\psi)+\ldots+{\cal L}_{\rm light}.
\end{eqnarray}
Loop graphs involve large momentum of order $m_Q$. The large momentum 
regions are accounted for by adapting the couplings $d_i(\Lambda)$ 
order by order in $\alpha_s(m_b)$. By adding more operators the full 
scattering amplitude can be reproduced to any accuracy in an 
expansion in $\alpha_s$ and $v$.

The Lagrangian (\ref{nrqcd}) appears 
similar to the heavy quark effective Lagrangian, except that it 
contains an antiquark clone of the single heavy quark sector and 
a quark-antiquark sector (operators made out of $\psi$ and $\chi$), 
which is absent in HQET. Indeed, the couplings in the single quark 
(antiquark) sector are identical in HQET and NRQCD to all orders 
in $\alpha_s$ \cite{Man97}, if the same factorization prescription 
is used. However, the power counting is generally different. For 
example, the kinetic energy term $\vec{\partial}^{\,2}/(2 m_Q)$ is expected 
to be of the same order as $i \partial_0$, because non-relativistic 
$Q\bar{Q}$ systems have momenta of order $m_Q v$ and 
energies of order $m_Q v^2$. The situation is actually more 
complicated. Because the kinetic term is part of the leading order 
Lagrangian, scattering amplitudes computed with NRQCD depend on 
$m_Q$, $v$ and the cut-off $\Lambda$ in a non-trivial way. If one 
chooses $\Lambda$ several times $m_Q v$, every Feynman diagram 
is a complicated function of $v$. This should be compared to HQET, 
where, choosing the cut-off several times $\Lambda_{\rm QCD}$, 
every Feynman integral is just a number. The dependence on $m_Q$ 
is fixed by the over-all power and coupling of the operator in the 
effective Lagrangian. There is no problem of principle with non-trivial 
dependence on $v$, if the Lagrangian is defined with a hard cut-off 
larger than $m_Q v$ and if the NRQCD quantities are computed 
non-perturbatively as in lattice NRQCD.

However, for perturbative calculations of perturbative $Q\bar{Q}$ 
systems this is inconvenient. First, one calculates too much, because 
one will need the NRQCD amplitude only to some accuracy in the $v$ 
expansion. The full $v$ dependence has a technical price. It is more 
difficult to compute an integral which is a function of $v$ than an 
integral which is just a number. Second, one would like to use 
dimensional regularization (DR). Here the subtleties arise, because 
DR gets NRQCD integrals, written down naively, wrong. Because the 
integrand depends on $m_Q$, the dependence on the scale $\mu$ of DR 
corresponds to a cut-off $\Lambda\gg m_Q$ instead of 
$m_Q\gg \Lambda\gg m_Q v$ \cite{Ben97}. Consequently, QCD is not 
matched correctly onto NRQCD. This difficulty is related to the 
existence of two scales, $m_Q v$ and $m_Q v^2$ in NRQCD. The first 
attempt to separate these scales was made in \cite{Lab96} in the 
context of cut-off regularizations and time-ordered perturbation 
theory in Coulomb gauge. This work introduced the important distinction 
of soft and ultrasoft gluons (photons) and the multipole expansion 
for ultrasoft gluons, but the construction remained complicated and 
somewhat qualitative. Subsequent work \cite{LM97,GR98,LS98} 
identified different momentum regions that should contribute to 
NRQCD integrals, but dropped the soft region identified 
in \cite{Lab96}. These early works introduced many of the important 
concepts that appear in a non-relativistic EFT construction,  
pointed towards the necessity to perform expansions of Feynman 
{\em integrands} \cite{Man97,Ben97,GR98} in order to define NRQCD in 
DR, but did not yet provide a complete solution to the problem 
of separating the scales $m_Q v$ and $m_Q v^2$ and of defining 
NRQCD in DR.

\subsection{Threshold expansion}

Take an on-shell scattering amplitude of heavy quarks with 
momentum $m_Q v$ and gluons with momentum $m_Q v^2$ in QCD. 
The basic problem is to construct {\em term by term} the expansion of 
this amplitude in $v$. Such a construction may introduce divergent 
loop integrals in intermediate steps, even if the original expression 
was finite. We would like to use DR to regularize these 
divergences and we require that any integral that we have to compute 
contributes only to a single order in the $v$ expansion. Solving this 
problem is equivalent to constructing a non-relativistic effective 
theory in DR with easy velocity power counting and all scales 
separated.

The expansion method described in \cite{BS98} begins by identifying 
the relevant momentum regions in loop integrals, which follow from 
the singularity structure of the Feynman integrand. This is analogous 
to identifying hard, collinear and soft particles in high-energy 
scattering of massless particles. For non-relativistic scattering of 
heavy quarks, one finds four momentum regions:
\begin{eqnarray}
\label{terminology}
\mbox{{\em hard} (h):} && l_0\sim \vec{l}\sim m_Q,\nonumber\\
\mbox{{\em soft} (s):} && l_0\sim \vec{l}\sim m_Q v,\nonumber\\
\mbox{{\em potential} (p):} && l_0\sim m_Q v^2,\,\,\vec{l}\sim m_Q v,
\nonumber\\ 
\mbox{{\em ultrasoft} (us):} &&  l_0\sim \vec{l}\sim m_Q v^2.
\end{eqnarray}
Both, heavy quarks and gluons can be hard, soft and potential, but only 
gluons can be ultrasoft. (In the following, ``gluons'' may include all 
other massless modes, i.e. light quarks and ghosts.)

The threshold expansion is constructed by writing a Feynman diagram 
in QCD as a sum of terms that follow from dividing each loop momentum 
integral into these four regions. The division is done implicitly, 
through expansion of the propagators. No explicit cut-offs are needed. 
The expansion rule is that in every region one performs a Taylor 
expansion in the quantities which are small in that region. An immediate 
consequence of this is that every integral contributes only 
to a single power in the velocity expansion.

To give an example of the expansion rules, consider the propagator 
of a heavy quark with momentum $(q/2+l_0,\vec{p}+\vec{l}\,)$, 
\begin{equation}
\frac{1}{l_0^2-\vec{l}^2-2\vec{p}\cdot\vec{l}-\vec{p}^{\,2}+q l_0+
y+i\epsilon},
\end{equation}
and assume that $\vec{p}$ scales as $m v$ and $y=q^2/4-m^2$ as $m v^2$. 
When $l$ is hard, we expand the terms involving $\vec{p}$ and $y$ 
and the leading term in the expansion scales 
as $v^0$. When $l$ is soft, the term $q l_0$ is largest and the remaining 
ones are expanded. The propagator becomes static and scales as $v^{-1}$. 
Notice that this means that the kinetic energy term in the NRQCD 
Lagrangian is treated as an interaction term in the soft region, because 
it is small. When $l$ is potential, the propagator takes its standard 
non-relativistic form after expansion of $l_0^2$ and scales as $v^{-2}$. 
The gluon propagator takes its 
usual form, when the gluon line is soft and ultrasoft and scales 
as $v^{-2}$ and $v^{-4}$, respectively. If the gluon momentum is 
potential, 
one can expand $l_0^2$ and the interaction becomes instantaneous. 
If we add the scaling rules for the loop integration measure, 
$d^4l\sim 1\,\mbox{(h)},\,v^4\,\mbox{(s)},\,v^5\,\mbox{(p)},\,v^8\mbox{(us)}$, 
we can immediately estimate the size of the leading term from a given 
region. It is clear that these rules can be reformulated as 
an effective Lagrangian. The hard subgraphs give the dimensionally 
regularized couplings in the NRQCD Lagrangian. On the other hand, 
the distinction of soft, potential and ultrasoft momentum implies 
a manipulation of the NRQCD Lagrangian that is not evident from 
(\ref{nrqcd}).

\begin{figure}[t]
   \vspace{-4.1cm}
   \epsfysize=24cm
   \epsfxsize=18cm
   \centerline{\epsffile{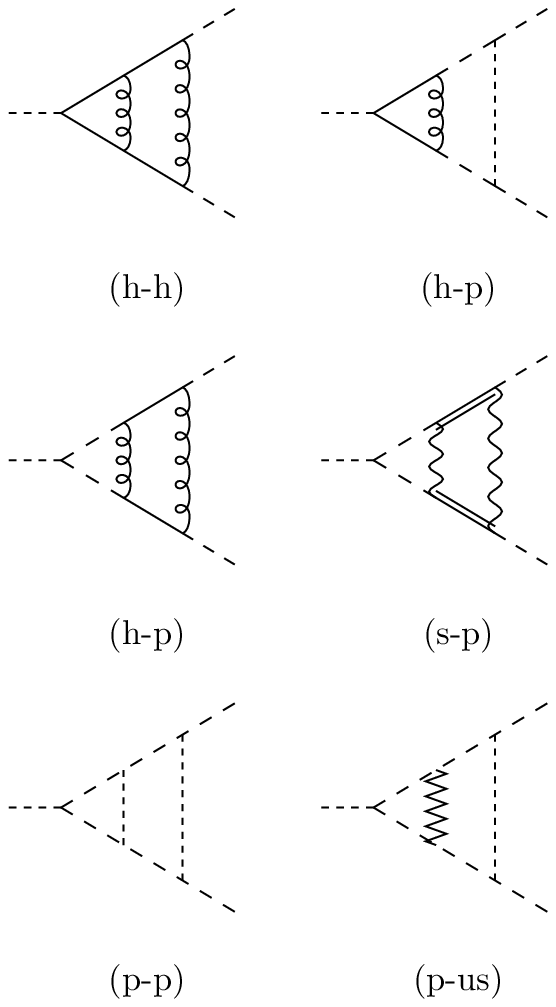}}
   \vspace*{-11.4cm}
\caption[dummy]{\label{fig1}\small 
Decomposition of the planar, 2-loop vertex integral in the threshold 
expansion. Line coding: solid and curled -- hard quarks and gluons, 
respectively; double line and wavy -- soft quarks and gluons; long- 
and short-dashed -- potential quarks and gluons; zigzag -- ultrasoft 
gluons.}
\end{figure}

It is instructive to discuss the physical interpretation of the terms 
that arise in the threshold expansion on a particular diagram. Take 
the planar, 2-loop correction to the electromagnetic 
heavy quark production vertex \cite{BS98}, i.e. the matrix element 
$\langle \bar{Q}(p_1) \bar{Q}(p_2)|\bar{Q}\gamma^\mu Q|0\rangle$. 
Many of the possible combinations of h, s, p and us loop momentum 
result in scaleless integrals, which are zero in dimensional 
regularization. Such scaleless integrals gain significance only in 
the context of the renormalization group, in which case one has to be 
more careful about the nature of $1/\epsilon$ poles. The non-vanishing 
configurations are shown in figure~\ref{fig1}. 
Every diagram stands for a series in $v$ that arises from the expansion 
of the integrand in a particular loop momentum configuration, but every 
integral in this series contributes only to a single power of $v$. 

The h-h configuration is a 2-loop correction to the coefficient 
functions in the non-relativistic expansion of the current 
$\bar{Q}\gamma^i Q \to C_1\psi^\dagger \sigma^i\chi +\ldots$. 
The leading term is of order $\alpha_s^2 v^0$. There 
are two possibilities for having 1-loop hard subgraphs. The 
first one (upper right) represents a 1-loop renormalization of 
the non-relativistic external current followed by exchange of 
a potential gluon. At leading order in the $v$ expansion potential 
gluon exchange can be interpreted as an interaction through the Coulomb 
potential $-\alpha_s C_F/r$. This contribution is of order 
$\alpha_s^2/v$. The second h-p term corresponds to the insertion 
of a four-fermion operator $(\psi^\dagger \kappa\psi)(\chi^\dagger\kappa
\chi)$ from the NRQCD Lagrangian. This contribution be\-gins at order 
$\alpha_s^2 v$. The soft subgraph in the s-p term can be interpreted as 
an instantaneous interaction, because a soft subgraph can be expanded 
in the zero components of its external momenta. The s-p graph 
corresponds to an insertion of (part of) the 1-loop corrected Coulomb 
potential plus higher potentials more singular in $r$. It contributes 
at order $\alpha_s^2/v$. The p-p term is dominant at small $v$. 
The double insertion of the Coulomb potential contributes at order 
$\alpha_s^2/v^2$. If $v$ is counted as $\alpha_s$, as forced upon us by 
the dynamics of perturbative, non-relativistic bound states, this 
term is unsuppressed relative to the tree graph. This shows the need 
to partially sum the expansion in $\alpha_s$ to all orders. Finally, 
ultrasoft gluon exchange is of order $\alpha_s^2/v$. This is actually an 
over-estimate. After combining all diagrams with an ultrasoft gluon, 
one finds that the coupling of ultrasoft gluons to heavy quarks has 
at least a factor of $v$. This cancellation is manifest in Coulomb 
gauge, in which only the spatial component of the gauge field can be 
ultrasoft. Hence the p-us term is of order $\alpha_s^2 v$. Note that 
in an expansion scheme in which $v\sim \alpha_s$, only four of the 
six terms in figure~\ref{fig1} are needed at next-to-next-to-leading 
(NNLO) order.

\subsection{Effective theory again: PNRQCD}

We have now defined NRQCD and can compute its couplings in DR. This 
leads to a new set of graphs in which hard subgraphs have been contracted 
to a point. The remaining loop integrals can still be soft, potential 
or ultrasoft. One could introduce different fields for p, s and us 
quarks and gluons and implement the threshold expansion rules at the 
level of effective propagators and vertices \cite{Gri98}. But since 
soft heavy quarks and gluons and potential gluons do not appear as 
external lines for non-relativistic systems, we can integrate them out. 
As can be seen from the structure of the threshold expansion, soft 
subgraphs have the same combinatorical structure as hard subgraphs. 
They can be contracted to effective operators, which are non-local 
in space, but local in time (instantaneous). These effective operators 
provide a generalized notion of the heavy quark potential. 
The resulting effective theory contains only potential quarks and 
ultrasoft gluons. Velocity power counting is then trivial. 
The scheme is as follows:
\begin{displaymath}
\begin{array}{cc}
{\cal L}_{\rm QCD}\,[Q(h,s,p),\,g(h,s,p,us)] \quad& \mu>m \\[0.1cm]
\downarrow & \\[0.1cm]
{\cal L}_{\rm NRQCD}\,[Q(s,p),\,g(s,p,us)] \quad&  m v < \mu < m \\[0.1cm]
\downarrow & \\[0.1cm]
{\cal L}_{\rm PNRQCD}\,[Q(p),\,g(us)] \quad
& \mu < m v\\
\end{array}
\end{displaymath}
Such a construction has been proposed first, by tree level matching, 
 in \cite{PS98}. I follow \cite{PS98} in calling the second EFT 
potential NRQCD 
(PNRQCD). In the context of the threshold expansion, which provides a 
matching prescription for loop graphs, PNRQCD has been discussed 
in \cite{Ben98a,BSS99}. A somewhat different, but probably conceptually 
equivalent construction has been proposed recently in \cite{LMR99}. 

In general the PNRQCD Lagrangian can be written as
\begin{equation}
\label{loc}
{\cal L}_{\rm PNRQCD}={\cal L}^\prime_{\rm NRQCD} + 
{\cal L}_{\rm non-local}.
\end{equation}
${\cal L}_{\rm non-local}$ collects all non-local interactions. 
The local interactions are exactly those of NRQCD, but the interpretation 
is different, because only potential heavy quarks and 
ultrasoft gluons are left over. 
In diagrams constructed from ${\cal L}^\prime_{\rm NRQCD}$ the 
gluon propagators are always expanded according to their ultrasoft 
scaling rule and the heavy quark propagator has the familiar 
non-relativistic form, while in diagrams constructed from 
${\cal L}_{\rm NRQCD}$  gluons are also soft and potential and the heavy 
quark propagator can also be static. 
Because only potential quarks and ultrasoft gluons are left in PNRQCD, 
the interaction terms have definite velocity scaling rules. 
A potential quark propagator in coordinate space scales as $v^3$, so 
a quark field in PNRQCD scales as $v^{3/2}$. An ultrasoft gluon field 
counts $v^2$. 

Integrating out potential gluon exchange between a quark and an 
antiquark at tree level gives the leading order Coulomb potential. The 
unperturbed PNRQCD Lagrangian is 
\begin{eqnarray}
\label{pnrqcd}
{\cal L}_{\rm PNRQCD}^0 &=& \\
&&\hspace*{-1.5cm}
\psi^\dagger \left(i \partial^0+
\frac{\vec{\partial}^{\,2}}{2 m_Q}
\right)\psi + 
\chi^\dagger \left(i \partial^0-\frac{\vec{\partial}^{\,2}}{2 m_Q}
\right)\chi 
\nonumber\\
&&\hspace*{-1.5cm}
+\int d^3\vec{r}\,\left[\psi^\dagger T^A\psi\right](\vec{r}\,)\,
\left(-\frac{\alpha_s}{r}\right)\left[\chi^\dagger T^A\chi\right](0).
\nonumber
\end{eqnarray}
Since $v\sim \alpha_s(m_Q v)$ is assumed, all terms scale as $v^5$; 
the Coulomb interaction cannot be treated as a perturbation as is 
of course anticipated. 
One can rewrite the PNRQCD Lagrangian 
in terms of a `tensor field' $[\psi\otimes
\chi^\dagger](t,\vec{R},\vec{r}\,)$ that depends on the cms and 
relative coordinates. The unperturbed Lagrangian describes free 
propagation (with mass $2 m$) in the cms coordinate. The propagation 
of $[\psi\otimes \chi^\dagger](t,\vec{R},\vec{r}\,)$ in its 
relative coordinate is given by the Coulomb Green function of a 
particle with reduced mass $m/2$. In calculating diagrams with Coulomb 
Green functions, one sums corrections of order $(\alpha_s/v)^n$ to 
all orders. The remaining terms can be treated as perturbations in 
$v$ and $\alpha_s$ around the unperturbed Lagrangian. 

Higher order non-local interactions follow after matching potential 
gluon exchange to better accuracy and after integrating out soft 
loops. We can match assuming $\alpha_s\ll v$, and treat the Coulomb 
potential as a perturbation when matching 
NRQCD on PNRQCD. A general non-local operator is a function of 
$r$ and derivatives $\partial_i$ acting on the (anti)quark field. 
It is non-polynomial in $r$, but polynomial in $\partial_i$. In 
general, it may have ultrasoft gluon fields attached to it. Non-local 
operators are singular, for example $\alpha_s\delta^{(3)}(\vec{r})$ or 
$\alpha_s^2/r^2$. These singularities are harmless, because PNRQCD 
is defined with a cut-off (we imply dimensional regularization) and 
more singular non-local operators are in fact suppressed in $v$. 
The dimensionally 
regulated quark-antiquark potential 
up to order $v^7$, projected on a colour singlet, spin-1 $Q\bar{Q}$ 
pair can be found in \cite{BSS99}. 

\begin{figure}[t]
   \vspace{-6.8cm}
   \epsfysize=24cm
   \epsfxsize=18cm
   \centerline{\epsffile{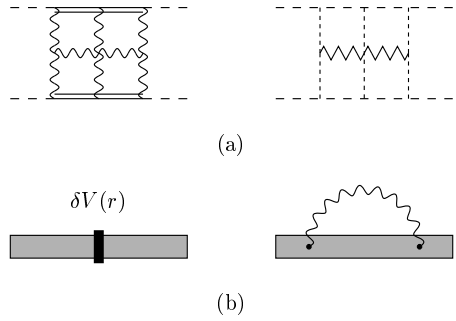}}
   \vspace*{-12.8cm}
\caption[dummy]{\label{fig2}\small (a) s-s-s region that gives 
rise to an infrared logarithm (left); p-p-us region which contains 
the corresponding ultraviolet logarithm. (b) In PNRQCD notation 
the two NRQCD graphs in (a) are interpreted as a (Coulomb) 
potential insertion (left) and an ultrasoft 1-loop diagram (right). 
The shaded bar represents the propagation of the $\bar{Q} Q$ 
according to the Coulomb Green function. Line coding as in 
figure~\ref{fig1}.}
\end{figure}
Potentials are short-distance 
coefficients, i.e. they are for PNRQCD what the 
$d_i$ in (\ref{nrqcd}) are for NRQCD. Consider the Coulomb potential 
in (\ref{pnrqcd}). Its coefficient is renormalized such that 
\begin{equation}
\alpha_s\,\to\, \alpha_s(\mu) v_c(\alpha_s,\mu r).
\end{equation}
The coefficient function $v_c(\alpha_s,\mu r)$ contains logarithms 
of $\mu r$, which are analogous to logarithms of the heavy quark 
mass in the coefficient functions of the NRQCD Lagrangian. 
Up to order $\alpha_s^2$, to which $v_c(\alpha_s,\mu r)$ is known 
exactly \cite{Sch99}, all logarithms $\ln \mu r$ can be absorbed 
into $\alpha_s(1/r)$. This is no longer true at three loops. 
There exists an uncancelled infrared divergence in the left 
diagram of figure~\ref{fig2}a \cite{ADM77} which, after subtraction 
of the pole in DR, gives rise to a logarithm not related to the 
running coupling. This logarithm is analogous to a non-vanishing 
anomalous dimension of local operators in the NRQCD Lagrangian. 
The scale dependence is cancelled for a physical process by the 
scale dependent PNRQCD matrix element, in this particular case 
the right diagram of figure~\ref{fig2}a. In PNRQCD notation, 
which does not make use of potential and soft lines, the correspondence 
is shown in figure~\ref{fig2}b. The interpretation of 
potentials as matching coefficients implies that the Coulomb potential 
is not identical to the static potential defined as the vacuum 
expectation of a Wilson loop in the limit $T\to\infty$ \cite{BPSV99}. 
The Coulomb potential is logarithmically sensitive to the 
ultrasoft scale, but it is not infrared divergent. This statement 
applies to any other potential. 

Although the local interactions in (\ref{loc}) 
are the same as those in NRQCD, we must 
write them in a different form to account for the expansion rules 
for ultrasoft gluons.
When an ultrasoft gluon line with momentum $l$ connects to a 
quark line with momentum $k-l/2$ for the incoming and $k+l/2$
for the outgoing quark line, the threshold expansion 
instructs us to expand the quark-gluon vertex 
and quark propagator in $\vec{l}/{\vec{k}}\sim v$. 
All gluon interaction 
terms in ${\cal L}^\prime_{\rm NRQCD}$ should be understood as 
multipole-expanded, for instance \cite{Lab96,GR98}
\begin{eqnarray}
&&\hspace*{-1cm}\left[\psi^\dagger A^i \partial^i\,
\psi\right](x) \equiv
\psi^\dagger(x)A^i(t,0)\partial^i\psi(x) 
\nonumber\\
&&\hspace*{-0.5cm}+\, 
\psi^\dagger(x)\,(x^j\partial^j A^i(t,0))
\partial^i \psi(x) + \ldots,
\end{eqnarray}
and likewise for all other terms in the NRQCD Lagrangian. 
The leading ultrasoft interactions 
follow from multipole expansion of the gauge field terms in 
$\psi^\dagger (i D^0+\vec{D}^{\,2}/(2 m_Q))\psi$ in (\ref{nrqcd})  
together with a non-abelian non-local term that comes from matching 
the graph in figure~\ref{fig3} on PNRQCD. The following collects all 
ultrasoft interactions up to order $v^{13/2}$: 
\begin{figure}[t]
   \vspace{-3.1cm}
   \epsfysize=20cm
   \epsfxsize=15cm
   \centerline{\epsffile{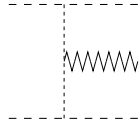}}
   \vspace*{-15cm}
\caption[dummy]{\label{fig3}\small 
NRQCD graph that generates the mixed non-local/ultrasoft interaction 
in (\ref{ultrasoft}). Line coding as in figure~\ref{fig1}.}
\end{figure}
\begin{eqnarray}
\label{ultrasoft}
{\cal L}_{us}&=&g_s[\psi^\dagger T^A\psi](x) \,A^{0A}(t,0)
\nonumber\\[0.2cm]
&&\hspace*{-0.8cm}
+\,g_s[\psi^\dagger T^A\psi](x)\,x^i\partial^i A^{0A}(t,0)
\nonumber\\[0.1cm]
&&\hspace*{-0.8cm}
-\,\frac{i g_s}{2 m_Q}\,[\psi^\dagger(\stackrel{\leftarrow}{\partial^i}
-\stackrel{\rightarrow}{\partial^i}) T^A\psi](x) \,A^{iA}(t,0)
\nonumber\\[0.1cm]
&&\hspace*{-0.8cm}+\,\mbox{antiquark terms}
\nonumber\\[0.1cm]
&&\hspace*{-0.8cm}
-\,\int d^3\vec{r}\,\left[\psi^\dagger T^B\psi\right](x+\vec{r}\,)\,
\left[\chi^\dagger T^C\chi\right](x)
\nonumber\\
&&\hspace*{-0.3cm}
\cdot \left(-\frac{\alpha_s}{r}\right) g_s f^{ABC} r^i A^{iA}(t,0)
\end{eqnarray}
The first line scales as $v^{1/2}$ relative to the leading order 
$v^5$ terms 
in the PNRQCD Lagrangian. The other three interaction terms scale as 
$v^{3/2}$. Using $[x^i,\vec{\partial}^{\,2}]=-2\partial^i$ and the 
equation of motion at leading order in $v$, which includes the 
Coulomb potential, the $v^{3/2}$ interactions combine into a 
chromo-electric dipole operator up to higher order terms. 
(Note that this shows that 
the distinction of non-local and local operators in PNRQCD is 
ambiguous, because they can be converted into each other by the 
equation of motion. Likewise a classification in powers of 
$1/m_Q$ is not useful.) Introducing 
the ultrasoft covariant derivative $D^0=\partial^0-i g_s A^0(t,0)$, 
we obtain 
\begin{eqnarray}
\label{pnrqcd2}
{\cal L}_{\rm PNRQCD} &=& \\
&&\hspace*{-1.8cm}
\psi^\dagger(x) \left(i D^0+
\frac{\vec{\partial}^2}{2 m_Q}-g_s x^i E^i(t,0)
\right)\psi(x) 
\nonumber\\[0.1cm]
&&\hspace*{-1.8cm}+\,\mbox{antiquark terms}
\nonumber\\[0.2cm]
&&\hspace*{-1.8cm}
+\,\int d^3\vec{r}\,\left[\psi_a^\dagger\psi_b\right]\!(x+\vec{r}\,)\,
V_{ab;cd}(r,\partial^i)\left[\chi_c^\dagger \chi_d\right]\!(x)
\nonumber
\end{eqnarray}
where
\begin{equation}
V_{ab;cd}(r,\partial^i)= 
T^A_{ab} T^A_{cd}
\cdot \left(-\frac{\alpha_s}{r}\right)
+\delta V_{ab,cd}(r,\partial^i)
\nonumber
\end{equation}
This Lagrangian is manifestly invariant under ultrasoft gauge 
transformations $U(t,0)$. Note that the spatial components of the gauge field 
transform covariantly under ultrasoft gauge transformations. Beyond 
tree-level the coefficient of the chromo-electric dipole operator 
receives corrections that can be computed in an expansion in $\alpha_s$. 
The PNRQCD Lagrangian (\ref{pnrqcd2}) appears to be equivalent to the 
Lagrangian derived in \cite{PS98}. The difference is that \cite{PS98} 
introduces a colour decomposition of the tensor field  $[\psi\otimes
\chi^\dagger](t,\vec{R},\vec{r}\,)$ and expresses the Lagrangian in 
terms of a colour singlet and a colour octet field $S$ and $O$. 

A Green function with no external ultrasoft lines requires at least 
two ultrasoft interactions. The leading ultrasoft correction is order 
$v$ from the ultrasoft covariant derivative in (\ref{pnrqcd2}). 
However, we can use 
an ultrasoft gauge transformation to gauge $A^0$ away, and hence 
the leading ultrasoft correction is order $v^3$. The Lagrangian 
(\ref{pnrqcd2}) can be used to compute all leading ultrasoft 
contributions. In particular, ultraviolet renormalization of ultrasoft 
graphs cancels the scale dependence of the potentials to order 
$v^3$. In \cite{BPSV99b} the scale dependence of the potentials 
has been computed using this correspondence. 

\subsection{Non-relativistic renormalization group}

The scale hierarchy $m_Q\gg m_Q v\gg m_Q v^2$ also implies large 
logarithmic corrections $\alpha_s\ln v$ and $\alpha_s\ln v^2$ to the 
coefficient functions. These logarithms can be summed to all orders in 
perturbation theory as follows:
\begin{itemize}
\item[1)] Match QCD and NRQCD at a scale $\mu_h\sim m_Q$, i.e. 
compute the coefficients $d_i$ as expansions in $\alpha_s(\mu_h)$. 
\item[2)] Compute the renormalization group scaling of the 
$d_i$ in NRQCD and evolve them to a scale 
$\mu_s\sim m_Q v\sim 1/r$.
\item[3)] Match NRQCD and PNRQCD at the scale $\mu_s$, i.e. 
compute the potential coefficients $v_i$ as expansions in 
$\alpha_s(\mu_s)$ and in terms of the $d_i(\mu_s)$.
\item[4)] Compute the renormalization group scaling of the 
potentials in PNRQCD and evolve them to a scale 
$\mu_{us}\sim m_Q v^2\sim m_Q\alpha_s^2$.
\item[5)] Compute the PNRQCD matrix elements with ultraviolet 
subtraction scale $\mu_{us}$.
\end{itemize}
I briefly discuss items 2) and 4), but note that an explicit 
calculation remains yet to be done.

NRQCD contains a single-heavy quark sector, which is identical to 
heavy quark effective theory (HQET). Since heavy quark-antiquark operators 
do not mix into this sector, its renormalization can be discussed 
separately. Operator renormalization in NRQCD arises from ultraviolet 
divergences in potential and soft loops. The single-heavy quark 
sector has no potential loops (all quark poles on one side of the 
real axis in the complex plane of loop momentum zero 
components); the anomalous dimension matrices are identical to those 
in HQET. It is convenient not to introduce non-local time-ordered 
product operators as is usually done in HQET, but to have lower 
dimensional operators mix into higher dimensional ones. For example, 
a vertex diagram with two insertions of the chromo-magnetic operator 
of (\ref{nrqcd}) requires a counterterm proportional to the Darwin 
interaction, hence 
\begin{equation}
\mu_1\frac{d}{d\mu_1} \,d_2 = -\frac{5\alpha_s}{2\pi}\,d_1^2+\ldots
\end{equation}
etc.. The single-heavy quark sector mixes into heavy quark-antiquark 
operators. These operators are renormalized by soft and potential 
loops. Mixing through potential loops is responsible for the scale 
dependence of the quark-antiquark current $\psi^\dagger \sigma^i 
\chi$ that appears first at two loops \cite{BSS98,CM98}. Higher dimension 
operators in the single-heavy quark sector can  
mix into lower dimension operators in the heavy quark-anti\-quark 
sector through potential 
loops, because potential gluon exchange can contribute factors 
$m_Q/|\vec{p}|$, where $\vec{p}$ is a relative momentum of order 
$m_Q v$. This never happens for soft loops. One can introduce two 
separate renormalization scales $\mu_1$ and $\mu_2$ for soft and 
potential loops and compute the corresponding anomalous dimension 
matrices. NRQCD coefficient functions $C_i(\mu_1,\mu_2)$ 
then depend on both scales. In the end we only need $C_i(\mu,\mu)$, 
which evolves with the sum of the two anomalous dimension matrices 
and the distinction is not necessary. However, the fact that potential 
loops are infrared finite tells us that the evolution in $\mu_2$ 
stops at a scale of order $m_Q v$. Soft loops are not always 
infrared finite; this connects the evolution in $\mu_1$ to the 
ultrasoft region. 

Identifying $\mu_1=\mu_2$ the NRQCD evolution terminates at the scale 
$\mu_s$ at which one matches to PNRQCD. The evolution of the 
potentials and other operators such as the chromo-electric dipole 
operator is then determined by the ultraviolet behaviour of ultrasoft 
loops.

A different implementation of renormalization group scaling from 
the one presented here has been suggested in \cite{LMR99}. In this 
case, too, an explicit calculation of operator renormalization 
has not yet been performed and the equivalence of the two approaches 
remains to be demonstrated.

\subsection{$\Lambda_{\rm QCD}\sim m_Q v^2$ or larger}
\label{larger}

The theoretical framework described so far requires $m_Q v^2\gg 
\Lambda_{\rm QCD}$. Nice as it may be, it can be applied safely only to 
extremely elusive systems such as toponium. There will be 
non-perturbative corrections suppressed by powers of 
$\Lambda_{\rm QCD}/(m_Q v^2)$ in addition to what I discussed 
above, which can be estimated by applying an operator product 
expansion to PNRQCD matrix elements \cite{Leu81}. In case of toponium 
these corrections are estimated to be small for the inclusive 
production cross section. Note that there are no 
non-perturbative modifications of the potentials, because they 
are short-distance objects.

What happens if $m_Q v^2$ is not that large?

If $m_Q\gg\Lambda_{\rm QCD}\,
\begin{array}{c}\sim\vspace{-21pt}\\[0.1cm]>\end{array}
\, m_Q v$, PNRQCD makes 
no sense. One can use NRQCD, but the NRQCD matrix elements are 
non-perturbative.

If $m_Q v\gg \Lambda_{\rm QCD}\,
\begin{array}{c}\sim\vspace{-21pt}\\[0.1cm]>\end{array}
\, m_Q v^2$, one 
can match perturbatively (in $\alpha_s(m_Q v)$) to PNRQCD, 
but since $\alpha_s(m v^2)\sim 1$ the self-coupling of ultrasoft 
gluons is unsuppressed. 
Velocity power counting is different from 
the one above which used $\alpha_s(m v^2)\sim \alpha_s(m_Q v)\sim v$. 
Non-perturbative gluons screen the ultrasoft scale $m_Q v^2$ and 
$\Lambda_{\rm QCD}$ takes its role. 
The coupling of ultrasoft gluons to heavy quarks is non-perturbative 
but small, of order $\Lambda_{\rm QCD}/(m_Q v)$. 
Hence ultrasoft effects enter a 
heavy quark-antiquark 
scattering amplitude as an uncalculable non-perturbative contribution 
beginning at order $(\Lambda_{\rm QCD}/(m_Q v))^2$. 
Up to this accuracy the amplitude is 
still determined by potentials. In particular, the potentials 
and the scale dependence of PNRQCD matrix elements remain perturbatively 
calculable.

\section{Heavy quark mass definitions}
\label{quarkmasses}

Before turning to applications of non-relativistic field theory, 
I discuss the concept of the heavy quark mass. The (P)NRQCD Lagrangian 
is normally expressed in terms of the quark pole mass and this 
has been assumed so far. If not I should have added a term 
$\delta m_Q\psi^\dagger\psi$ to the (P)NRQCD Lagrangian. There are 
good reasons to make use of this option, related to the fact that 
the pole mass, though an infrared safe quantity in perturbation theory 
\cite{Kro98}, incorporates uncalculable long-distance contributions 
of order $\Lambda_{\rm QCD}$. Originally discovered and discussed 
in the context of HQET \cite{BU94,BB94,BSUV94}, the problem is acute 
whenever heavy quarks are not off-shell by an amount of order $m_Q^2$. 
Recently there has been renewed interest in this problem in the 
context of $Q\bar{Q}$ systems \cite{Ben98,HSSW98} and suitable 
alternative definitions of the heavy mass have been proposed and 
used in applications. Such masses have the generic property that they 
differ from the pole mass by an amount linear in a subtraction 
scale $\mu_f$ \cite{BSUV94}. In this section I review the currently 
used mass definitions.

\subsection{$\overline{\rm MS}$ and pole mass}

The $\overline{\rm MS}$ mass $\overline{m}_Q(\mu)$ 
is the coefficient of the operator $\overline{Q} Q$ in QCD subtracted in the 
$\overline{\rm MS}$ scheme. It is best understood as a 
coupling constant just as $\alpha_s$. The $\overline{\rm MS}$ mass 
is scale-dependent. The scale dependence is related to loop momenta 
$l\gg m_Q$. For this reason, it makes no sense to evolve 
the $\overline{\rm MS}$ mass to scales parametrically smaller than 
$m_Q$. While formally possible, this generates fake logarithms of the 
ratio of scales. However, 
$\overline{m}_Q\equiv \overline{m}_Q(\overline{m}_Q)$ 
is a very useful reference parameter, just as $\alpha_s(m_Z)$.

The pole mass is the location of the pole in the full heavy quark 
propagator. (The weak interactions are switched off, so that quarks are 
stable. A finite decay width would not alter the conclusion of this 
subsection \cite{SW97}.) As such it is defined order by order 
in perturbation 
theory. Its relation to the $\overline{\rm MS}$ mass is given by 
\begin{equation}
\frac{m_Q}{\overline{m}_Q} = 1+\sum_{n=1} k_n\alpha_s(\overline{m}_Q)^n.
\end{equation}
For $b$ quarks, neglecting internal the charm quark mass effects, 
\begin{eqnarray} 
\label{polecoeffs}
&&\hspace*{-0.5cm}
k_1=0.424,\quad k_2=0.940,
\nonumber\\
&&\hspace*{-0.5cm}
k_3=3.096, \quad k_4=13.60
\end{eqnarray}
The first two coefficients are known analytically \cite{GBGS90}, the third 
is known numerically \cite{CS99}. For the fourth order coefficient I 
have used the so-called ``large-$\beta_0$'' estimate \cite{BB95}, which 
turned out to approximate $k_{2,3}$ very well.

Neither of the two masses is a useful parameter for perturbative 
calculations of $Q\bar{Q}$ systems. The $\overline{\rm MS}$ mass would 
imply $\delta m_Q\sim m_Q\alpha_s$ and hence $\delta m_Q\psi^\dagger\psi
\sim v^4$ would dominate the (P)NRQCD Lagrangian. One must have 
$\delta m_Q\sim m_Q v^2$ or smaller. The pole mass is long-distance 
sensitive at order $\Lambda_{\rm QCD}$. There would be nothing wrong with 
this if not for the following two facts: first, the quantities we 
would like to compute with non-relativistic field theory such as 
correlation functions of two heavy quark currents near threshold 
are less long-distance sensitive than the pole mass. This desirable property 
is lost, if one uses renormalization conventions which are more 
long-distance sensitive than the quantity of interest. Second, 
long-distance sensitivity is related to large perturbative 
corrections through infrared renormalons \cite{Ben99}. Here  
this means that series expansions diverge as $(2\beta_0)^n n! n^b$, where 
$\beta_0$ is defined through 
$d\alpha_s/d\ln\mu^2=-\beta_0\alpha_s^2+\ldots$. The divergence 
enters less long-distance sensitive quantities only through 
$k_{n+1}\sim (2\beta_0)^n n! n^b$ and would be much milder, if a 
suitable mass renormalization convention were implied.  For the following 
discussion it is useful to introduce an ``asymptotic counting'' for 
perturbative coefficients. The asymptotic counting of the coefficients 
(\ref{polecoeffs}) is $k_n\sim (n-1)!\,\mu/\overline{m}_Q$, where the 
definition of $k_n$ has been generalized to an expansion in 
$\alpha_s(\mu)$ and we neglect the factors $(2\beta_0)^n$ and $n^b$ 
in this schematic notation. I will also call a series expansion 
``convergent'', if it diverges less rapidly than the $k_n$. 
Coefficients of ``convergent'' series count as order 1 and terms 
in a convergent series are counted only according to their power in
$\alpha_s$. 

Asymptotic counting may appear abstract and irrelevant to 
next-to-next-to-leading order calculations. The physical systems 
which we discuss later (and numerous quantities related to $B$ meson 
decays) show that this is not so. This has led 
to heavy quark mass definitions, which satisfy $\delta m_Q \sim 
m_Q v^2$ and are convergent in asymptotic counting.

\subsection{PS mass}

The potential subtraction (PS) scheme \cite{Ben98} is based on 
the observation that there is a cancellation of divergent series 
behaviour in the combination $2 m_Q+[V(r)]_{\rm Coulomb}$. This 
can be seen explicitly at 1-loop and in the large-$\beta_0$ 
approximation \cite{Ben98,HSSW98} by combining the results of 
\cite{BB95,AL95} and by a diagrammatic argument at two loops 
\cite{Ben98} and, probably, in higher orders. The potential subtracted 
(PS) mass at subtraction scale $\mu_f$ is defined by 
\begin{eqnarray}
m_{Q,\rm PS}(\mu_f) &=& m_Q+\frac{1}{2}\!\int\limits_{|\vec{q}\,|<\mu_f}
\!\!\!\!\frac{d^3 \vec{q}}{(2\pi)^3}\,[\tilde{V}(q)]_{\rm Coulomb}
\nonumber\\[0.1cm]
&&\hspace*{-1.5cm}\equiv \,m_Q - \mu_f\sum_{n=0} l_n(\mu_f/\mu)\,
\alpha_s(\mu)^{n+1},
\end{eqnarray}
where 
\begin{equation}
[\tilde{V}(q)]_{\rm Coulomb} = -\frac{4\pi\alpha_s}{\vec{q}^{\,2}}\,
\tilde{v}_c(\alpha_s,q/\mu)
\end{equation}
is the Coulomb potential in momentum space defined as a PNRQCD 
coefficient function as discussed above. The coefficients 
$l_{0,1,2}$ are given in \cite{Ben98}. The large-$\beta_0$ 
estimate is easily derived from \cite{AL95} and I will use the 
result for $l_3$ below. Because the Coulomb potential is scale-dependent 
at order $\alpha_s^4$ and beyond, it depends on the PNRQCD matching 
scale. The coefficients $l_n$, $n>2$, inherit this scale, which has to 
be specified in addition to $\mu_f$. We can choose this scale equal 
to $\mu_f$, so that the logarithm in the Coulomb potential 
\cite{BPSV99} does not contribute to the PS mass.

In asymptotic counting $l_n\sim n!\,\mu/\mu_f$, so that the combination  
$m_Q k_n-\mu_f l_{n-1}$, which appears in the relation of $m_{Q,\rm PS}$ 
to $\overline{m}_Q$, is convergent as desired. This is true for 
any $\mu_f>\mbox{few}\times\Lambda_{\rm QCD}$. To satisfy $\delta m_Q
\sim m_Q v^2$ or smaller, $\mu_f$ must not be parametrically larger than 
$m_Q v$. It is useful to choose $\mu_f$ of order $m_Q v\sim m_Q\alpha_s$. 
Note that this implies that $m_Q k_n$ and $\mu_f l_{n-1}$ are of different 
order in $\alpha_s$, but cancel asymptotically. 

\subsection{1S mass}

The 1S scheme was proposed in \cite{HLM99} and uses one half of 
the perturbative $\Upsilon(1S)$ mass as quark mass parameter. 
The perturbative $\Upsilon(1S)$ mass is related to the physical 
$\Upsilon(1S)$ mass $M_{\Upsilon(1S)}$ by 
\begin{equation}
M_{\Upsilon(1S)}=2 m_{Q,1S}+\bar{\Lambda}_{\Upsilon}, 
\end{equation}
where $\bar{\Lambda}_{\Upsilon}$ is a poorly known non-perturbative 
contribution, which is most likely less (if not considerably less) 
than $100$--$150\,$MeV. (Some versions of the 1S scheme eliminate 
the bottom quark mass in favour of the physical $\Upsilon(1S)$ mass, 
the advantage being that the input parameter is a physical 
quantity which is very accurately measured. I prefer the version 
stated above, because it does not obscure the presence of 
an unknown non-perturbative contribution. The 1S scheme can also be 
defined for top quarks \cite{HT99}. It uses the perturbative toponium 1S 
mass under the assumption of a stable top.) Parametrically  
$\bar{\Lambda}_{\Upsilon}$ is of order 
$(\Lambda_{\rm QCD}/(m_Q v))^4$ and not of order $\Lambda_{\rm QCD}$. 
This guarantees that $m_{Q,1S}$ is less long-distance sensitive than 
the pole mass for perturbative $Q\bar{Q}$ systems.

The 1S scheme does not have an explicit subtraction scale $\mu_f$, 
but it is similar (up to ``finite'' renormalizations) to the PS scheme 
with $\mu_f$ of order $m_Q \alpha_s$. The series expansion is 
\begin{equation}
\label{1smass}
m_{Q,1S}/m_Q = 1-\alpha_s\sum_{n=0} e_n(\mu)\,\alpha_s(\mu)^{n+1}.
\end{equation}
The coefficients are known exactly up to $e_2$ \cite{PY98} and 
will be given in Sect.~\ref{masses} below. The asymptotic counting 
is $e_n\sim n!\,\mu/(m_Q\alpha_s)$, where the factor $\mu/(m_Q\alpha_s)$ 
follows from the fact that the physical scale is of order 
$m_Q\alpha_s$. Hence the combination $k_n-e_{n-1}\alpha_s$, which 
enters the relation of $m_{Q,1S}$ and $\overline{m}_Q$ as coefficient 
at order $\alpha_s^{n}$ forms a convergent series with coefficients 
of order 1. As expected from the correspondence to 
$\mu_f\sim m_Q\alpha_s$, one has to combine coefficients of different 
order in $\alpha_s$ \cite{HLM99}.

Strictly speaking the 1S scheme cannot be 
consistently used in NNLO (that is, keeping $e_2\alpha_s^4$ in 
(\ref{1smass})) and beyond for bottom quarks, since  
the $\Upsilon(1S)$ ultrasoft scale 
$m_Q\alpha_s^2$ is of 
order $\Lambda_{\rm QCD}$. As discussed in 
Sect.~\ref{larger}, in this case the leading ultrasoft contribution 
(which would be order $\alpha_s^5$ for a perturbative system) 
is non-perturbative and of the same parametric order as the term 
$e_2\alpha_s^4$. This should be kept in mind, but in the following 
I use (\ref{1smass}) as a formal definition of the scheme, 
including the NNLO term $e_2$.

\subsection{Kinetic mass}

The so-called kinetic mass has its roots in $B$ physics \cite{BSU97}. 
The $B$ meson mass has the heavy quark expansion
\begin{equation}
m_B = m_b+\bar{\Lambda}+\frac{\mu_\pi^2-\mu_G^2}{2 m_b}+\ldots,
\end{equation}
with $\mu_\pi^2$ and $\mu_G^2$ related to the matrix elements of the 
kinetic energy and chromo-magnetic operators, respectively. The kinetic 
mass can be understood as a perturbative evaluation of this formula, 
in which the matrix elements include loop momentum integration 
regions below the scale $\mu_f$:
\begin{eqnarray}
m_Q &=& m_{Q,\rm kin}(\mu_f)+[\bar{\Lambda}(\mu_f)]_{\rm pert}
\nonumber\\[0.1cm]
&&\hspace*{-0.0cm}+\,
\left[\frac{\mu_\pi^2(\mu_f)}{2 m_Q}\right]_{\rm pert}+\ldots.
\end{eqnarray}
The matrix elements on the right hand side subtract the long-distance 
sensitive contributions to the pole mass order by order in 
$\mu_f/m_Q$ and $\alpha_s$. While easily stated, the definition of 
power divergent matrix elements in perturbation theory is largely 
arbitrary. The convention for the kinetic mass used in the literature 
uses an indirect definition through heavy flavour sum rules \cite{BSU97}, 
which is rather complicated when compared with the other two mass 
definitions above. The relation between the kinetic and the pole mass 
is known exactly at order $\alpha_s^2$ (including terms of order 
$\mu_f^2/m_Q$) and in the large-$\beta_0$ limit at order $\alpha_s^3$ 
\cite{CMU97}.

\subsection{Comparison}
\TABULAR{|c|c|c|c|c|}{
\hline 
&&&&\\[-0.3cm]
\tt $\overline{m}_b$ & $m_{b,\rm PS}(2\,\mbox{GeV})$ & 
$m_{b,\rm kin}(1\,\mbox{GeV})$ & $m_{b,1S}$ &
$m_{b,\rm pole}$\\[0.1cm]
\hline
\multicolumn{5}{|c|}{{\small $\alpha_s(m_Z)=0.118$ 
[$\alpha_s(4.25\,\mbox{GeV})=0.2240$]}}\\
\hline
{\small 4.15} & {\small 4.36/4.44/4.47/{\sl 4.48}} & 
{\small 4.41/4.49/{\sl 4.50}/-} & 
{\small 4.36(50)/4.60(62)/4.67(66)/{\sl 4.66}(-)} & 
{\small 4.55/4.75/4.89/{\sl 5.04}}\\
\hline
{\small 4.20} & {\small 4.41/4.49/4.52/{\sl 4.54}} & 
{\small 4.46/4.54/{\sl 4.56}/-} & 
{\small 4.41(55)/4.66(68)/4.72(72)/{\sl 4.71}(-)} & 
{\small 4.60/4.80/4.95/{\sl 5.09}}\\
\hline
{\small 4.25} & {\small 4.46/4.55/4.58/{\sl 4.59}} & 
{\small 4.52/4.60/{\sl 4.61}/-} & 
{\small 4.46(61)/4.71(73)/4.78(77)/{\sl 4.76}(-)} & 
{\small 4.65/4.85/5.00/{\sl 5.15}}\\
\hline
{\small 4.30} & {\small 4.52/4.60/4.64/{\sl 4.65}} & 
{\small 4.57/4.65/{\sl 4.67}/-} & 
{\small 4.52(66)/4.76(78)/4.83(82)/{\sl 4.82}(-)} & 
{\small 4.71/4.91/5.06/{\sl 5.20}}\\
\hline
{\small 4.35} & {\small 4.57/4.66/4.69/{\sl 4.70}} & 
{\small 4.62/4.71/{\sl 4.72}/-} & 
{\small 4.57(71)/4.82(84)/4.88(88)/{\sl 4.87}(-)} & 
{\small 4.76/4.96/5.11/{\sl 5.25}}\\
\hline
\multicolumn{5}{|c|}{{\small $\alpha_s(m_Z)=0.121$ 
[$\alpha_s(4.25\,\mbox{GeV})=0.2355$]}}\\
\hline
{\small 4.15} & {\small 4.37/4.45/4.49/{\sl 4.51}} & 
{\small 4.42/4.51/{\sl 4.52}/-} & 
{\small 4.37(52)/4.63(65)/4.70(69)/{\sl 4.68}(-)} & 
{\small 4.57/4.79/4.96/{\sl 5.14}}\\
\hline
{\small 4.20} & {\small 4.42/4.51/4.55/{\sl 4.56}} & 
{\small 4.48/4.56/{\sl 4.58}/-} & 
{\small 4.42(57)/4.68(70)/4.75(75)/{\sl 4.73}(-)} & 
{\small 4.62/4.84/5.01/{\sl 5.19}}\\
\hline
{\small 4.25} & {\small 4.47/4.56/4.61/{\sl 4.62}} & 
{\small 4.53/4.62/{\sl 4.64}/-} & 
{\small 4.47(62)/4.73(76)/4.81(80)/{\sl 4.79}(-)} & 
{\small 4.67/4.90/5.07/{\sl 5.25}}\\
\hline
{\small 4.30} & {\small 4.53/4.62/4.66/{\sl 4.67}} & 
{\small 4.58/4.67/{\sl 4.69}/-} & 
{\small 4.53(68)/4.79(81)/4.86(85)/{\sl 4.84}(-)} & 
{\small 4.73/4.95/5.12/{\sl 5.30}}\\
\hline
{\small 4.35} & {\small 4.58/4.68/4.72/{\sl 4.73}} & 
{\small 4.64/4.73/{\sl 4.75}/-} & 
{\small 4.58(73)/4.84(86)/4.92(91)/{\sl 4.89}(-)} & 
{\small 4.78/5.00/5.18/{\sl 5.35}}\\
\hline}
{Comparison of $b$ quark masses for given $\overline{\rm MS}$ mass 
$\overline{m}_b$ at the scale $\overline{m}_b$ for two values of 
$\alpha_s(m_Z)$. We used $n_f=4$ and put $m_c=0$. Slanted 
numbers make use of large-$\beta_0$ estimates. All numbers in GeV.
\label{tab1}}

In table~\ref{tab1} I compare the various mass definitions for 
$b$ quarks numerically using the $\overline{\rm MS}$ mass as a 
reference parameter. Each entry gives the value of the mass using 
a 1-loop/2-loop/3-loop/4-loop relation. For reasons that will 
become clear later it is interesting to have four-loop accuracy. 
Where available I have used large-$\beta_0$ estimates to obtain 
the four-loop value. I used $\alpha_s(\overline{m}_b)$ as 
perturbative expansion parameter (with one exception, see below). In defining 
the kinetic mass, terms of order $(\mu_f/m_b)^3$ or less are 
dropped. At order $\alpha_s^3$ the large-$\beta_0$ estimate is used, 
dropping all known terms that are subleading in this limit. (The 
relevant formula is given in the preprint version of \cite{CMU97}.) 

The 1S mass is computed in two different ways. First, I express 
it as a series in $\alpha_s(\overline{m}_b)$ with coefficients 
$k_n-e_{n-1}\alpha_s(\overline{m}_b)$ as explained above. A 4-loop relation 
then requires $e_3$, which is not yet known, or at least its 
large-$\beta_0$ value. This result is shown in brackets in 
table~\ref{tab1}. The second way first computes the PS mass at the 
scale $\mu_f=2\,$GeV as series in $\alpha_s(\overline{m}_b)$ and then 
relates the 1S mass to the PS mass as an expansion in 
$\alpha_s(2\,\mbox{GeV})$. In this case $e_3$ is not needed. The result 
is shown without brackets in table~\ref{tab1}. The two ways of 
computing $m_{b,1S}$ correspond to the mass analyses performed in 
\cite{Hoa99} and \cite{BS99}, respectively. 

Except for the $b$ quark pole mass all other masses given in the table 
are related to the $\overline{\rm MS}$ mass by ``convergent'' series. 
The difference is clearly visible in the magnitude of successive 
perturbative terms. For the pole mass the minimal term is reached around 
3-4 loops suggesting that its best accuracy cannot be smaller 
than about $150\,$MeV. On the contrary, the perturbative expansions of 
the other masses behave extremely well. At fourth order the error in 
relating them to $\overline{m}_b$ is of order $10\,$MeV! 

At this level, 
a comment on the effect of keeping the mass of internal charm quarks is 
necessary. Charm quark mass effects at order $\alpha_s^2$ are known for 
the pole quark mass \cite{GBGS90}. For given $\overline{m}_b=4.25\,$GeV 
and $\overline{m}_c=(1.1-1.4)\,$GeV the pole mass increases by about 
$8$--$10\,$MeV. 
The charm mass effect on the PS and 1S mass is easily computed 
in terms of the charm quark contribution to the 
photon vacuum polarization. I find, again for given 
$\overline{m}_b$, that $m_{b,\rm PS}(2\,\mbox{GeV})$ and $m_{b,1S}$ 
are reduced by $11$--$12\,$MeV and $4$--$5\,$ MeV respectively. 
These corrections could be applied to table~\ref{tab1}, which is generated 
for $m_c=0$. Charm effects at order $\alpha_s^3$ and beyond are not 
known and introduce an error of perhaps $10\,$MeV. If the average 
internal loop momentum becomes small, the charm quark decouples and 
one can switch to an effective description with only three light 
flavours \cite{BB95}.

\subsection{Application: inclusive heavy quark decay}

Before returning to $Q\bar{Q}$ systems it is interesting to see how 
these mass definitions fare in decays of $B$ mesons. As an example 
consider the inclusive semi-leptonic $B$ decay $B\to X_u l\nu$. Many 
more examples can be found in \cite{HLM99}, though restricted to 
the 1S scheme. (An earlier 1-loop analysis with the kinetic mass was 
done in \cite{ura95}.)

It is known that inclusive $B$ decays are less long-distance sensitive 
than the pole mass \cite{BSUV94,BBZ94} and so another mass parameter 
is warranted. But the situation is different from onium systems, because 
there is no scale other than $m_b$. Hence  
$\overline{m}_b(\overline{m_b})$ is a legitimate choice. 
However, this works well only asymptotically but fails in low orders in 
$\alpha_s$ \cite{BBB95b}. The decay rate, neglecting power 
corrections, is
\begin{equation}
\Gamma = \frac{G_F^2|V_{ub}|^2 M^5}{192\pi^3}
\Big[1-\delta_1-\delta_2+O(\alpha_s^3)\Big].
\end{equation}
With the second order correction in the pole mass scheme from 
\cite{Rit99}, the 1-loop and 2-loop correction in various other 
mass renormalization conventions is computed and 
shown in table~\ref{tab2}. All 
``alternative'' mass definitions introduced above do very well.
 
\TABULAR{|c|c|c|}{
\hline 
&&\\[-0.3cm]
$M$ & $\delta_1$ & $\delta_2$ \\[0.1cm]
\hline
$m_{b,\rm pole}$ & $+0.17$ & $+0.10$ \\
\hline
$\overline{m}_b$ & $-0.30$ & $-0.13$ \\
\hline
$m_{b,\rm PS}(2\,\mbox{GeV})$ & $-0.03$ & $-0.01$ \\
\hline
$m_{b,1S}$ & $+0.11$ & $+0.03$ \\
\hline
$m_{b,\rm kin}(1\,\mbox{GeV})$ & $+0.03$ & $-0.005$ \\
\hline}
{\label{tab2} First and second order perturbative terms in inclusive 
$b\to u$ decay with various mass parameters.}

It is perhaps not evident why this is the case given that there 
is no a priori reason not to use the $\overline{\rm MS}$ mass. 
A plausible argument is that while the characteristic scale in 
inclusive $b\to u$ decay is parametrically of order $m_b$, it is 
numerically smaller \cite{BSUV97}. This forces the $b$ quark closer 
to its mass shell although not close enough to justify the use of the 
pole mass. Similar improvements by using ``alternative'' mass renormalization 
schemes occur for other heavy quark decays \cite{HLM99}. 

\section{Quarkonium}
\label{quarkonium}

In this section I review results on quarkonium bound states. 
Unfortunately, none of the observed charmonium and bottomonium 
states is truly perturbative, i.e. satisfies 
$m_Q v^2\gg \Lambda_{\rm QCD}$. The best candidate to try our 
theory are the $\Upsilon(1S)$ and $\eta_b$ states. Higher excitations 
are almost certainly non-perturbative, although more 
non-relati\-vi\-stic. 

\subsection{Masses}
\label{masses}

The quarkonium binding energy is of order $m_Q\alpha_s^2$ in leading 
order. For an arbitrary $Q\bar{Q}[nl]$ state the energy is known 
to order $\alpha_s^4$ \cite{PY98}. For $nS$ states the result of 
\cite{PY98} has been confirmed by \cite{PP98,MY98,BSS99}. The 
$\Upsilon(1S)$ mass, expressed in terms of the $b$ quark pole mass, 
is given by
\begin{eqnarray}
\label{ups1s}
M_{\Upsilon(1S)} &=& 2 m_b-\frac{4}{9}\,m_b^2\alpha_s^2 \,\bigg[
\,1+\frac{\alpha_s}{\pi}\bigg(-\frac{25}{6}\,l\nonumber\\
&&\hspace*{-1cm}
+\frac{203}{18}\bigg)
+\frac{\alpha_s^2}{\pi^2}\bigg(\frac{625}{48}\,l^2
-\frac{1429}{24}\,l-\frac{9\pi^4}{32}\nonumber\\
&&\hspace*{-1cm}
+\,\frac{2453\pi^2}{216}+
\frac{1235\zeta(3)}{36}+\frac{7211}{108}\bigg)\bigg]
\nonumber\\
&&\hspace*{-1.3cm}
=\,2 m_b-\frac{4}{9}\,m_b^2\alpha_s^2 \,\Big[1+1.08+1.76\Big],
\end{eqnarray}
where $l=\ln(16 m_b^2\alpha_s^2/(9\mu^2))$ and $\alpha_s=
\alpha_s(\mu)$ and the second line is given for $\mu$ such that 
$l=0$ (for $m_b=5\,$GeV), in which case $\alpha_s(\mu)=0.30$.  
Corrections to this result are order $\alpha_s^5$ and order 
\begin{equation}
\label{nonpert}
\alpha_s^2\left(\frac{\Lambda_{\rm QCD}}{m_b \alpha_s}\right)^2
\left(\frac{\Lambda_{\rm QCD}}{m_b\alpha_s^2}\right)^{2+2 n}
\end{equation}
from the operator product expansion \cite{Leu81}.

An interesting application of (\ref{ups1s}) is to use it to determine 
the $b$ quark mass. The non-\-convergence of the series (\ref{ups1s}) 
seems to make this impossible. However, if the series is expressed in 
terms of the PS mass or the $\overline{\rm MS}$ mass, the convergence 
is dramatically improved as seen from the column referring to 
$m_{b,1S}$ in table~\ref{tab1}. In \cite{BS99} this has been 
used to obtain
\begin{equation}
\label{mass1}
\overline{m}_b(\overline{m}_b) = (4.24\pm 0.09)\,\mbox{GeV}.
\end{equation}
The central value is obtained by varying $\mu$ from $1.25\,$GeV to 
$4\,$GeV and by symmetrizing the error. 
The total error is dominated by the unknown 
non-perturbative contribution from ultrasoft gluons and the fact that 
the OPE series in $n$ of (\ref{nonpert}) may not converge. 
Note that (\ref{mass1}) 
uses the 4-loop relation in table~\ref{tab1} because the series 
(\ref{ups1s}) in terms of the PS or $\overline{\rm MS}$ mass is 
convergent in asymptotic counting and hence determines the quark mass 
to order $\alpha_s^4$. This is the main reason why (\ref{mass1}) differs 
substantially from the value obtained in \cite{PY98}, 
where $\overline{m}_b$ is determined via the $b$ pole 
mass and by a 2-loop relation.

There exist partial results at order $\alpha_s^5$ which may be used 
to estimate the perturbative error on $M_{\Upsilon(1S)}$. The mass 
correction from ultrasoft gluon exchange has been obtained in 
\cite{KP99a}:
\begin{eqnarray}
\label{usest}
\delta M^{us}_{\Upsilon(1S)} &=& 6.31\,m_b\alpha_s^5
\left[\ln\left(\frac{9\mu_{us}}{8 m_b\alpha_s^2}\right)-2.06\right]
\nonumber\\[0.1cm]
&&\hspace*{-1.0cm}\approx\, -(35-250)\,\mbox{MeV}.
\end{eqnarray}
To obtain the estimate for the ultrasoft constant terms, I interpreted 
$\alpha_s^5$ as $\alpha_s(\mu_1)^4\alpha_s(\mu_{us})$, 
with $\mu_1\approx 2\,$GeV 
the scale that makes the logarithm $l$ in (\ref{ups1s}) vanish, and 
varied $\mu_{us}$ from $0.7\,$GeV to $2\,$GeV. The numerical 
estimate is highly sensitive to the PNRQCD cut-off scale $\mu_{us}$ 
which must cancel in a complete order $\alpha_s^5$ calculation. 

The second estimate is based on the logarithmically enhanced terms 
of order $\alpha_s^5\ln\alpha_s$ \cite{BPSV99b,KP99b}:
\begin{eqnarray}
\label{usest2}
\delta M^{ln}_{\Upsilon(1S)} &=& \frac{1730}{81\pi}\,
m_b\alpha_s^5 \ln(1/\alpha_s)
\nonumber\\[0.1cm]
&& \hspace*{-1.0cm}\approx\,(75-100)\,\mbox{MeV}.
\end{eqnarray}
Here I varied the scale in $\alpha_s\ln(1/\alpha_s)$ from $0.7\,$ 
to $2\,$GeV as above. This set of terms does not depend on arbitrary 
cut-offs, but the logarithm is not large. Note that the coefficient of 
the logarithm is not identical to the one in (\ref{usest}), 
because there are logarithms unrelated to ultrasoft effects in the 
potentials of PNRQCD. 

Both corrections may not be small compared to the error estimate in 
(\ref{mass1}), but since they come with opposite sign and constitute 
only part of the $\alpha_s^5$ correction it is too early to revise 
(\ref{mass1}). However, they illustrate that next-to-next-to-next-to-leading 
order effects may be large.

\subsection{Decay into a lepton pair}

The decay of a $nS$ state into $l^+ l^-$ measures the quarkonium 
wave function $\Psi(0)$ at the origin. In turn, this parameter 
enters all those quarkonium decays which proceed through 
$Q\bar{Q}$ annihilation. In terms of the quark 
electric charge $e_Q$, the fine structure constant $\alpha$ and the 
mass $M_{nS}$ of the quarkonium, the decay rate reads, for massless 
leptons:
\begin{equation}
\label{decay1}
\Gamma = \frac{16 \pi e_Q^2\alpha^2}{M_{nS}^2}\,C(\alpha_s;\mu)^2\,
|\Psi(0)|^2(\mu).
\end{equation}
Here $C(\alpha_s;\mu)$ is the short-distance 
coefficient of the non-relativistic vector current 
$\psi^\dagger\sigma^i\chi$, which is known to NNLO \cite{BSS98,CM98}, 
and $\Psi(0)$ is related to the NRQCD matrix element of the 
current. Note that the wave function at the origin is factorization 
scale dependent at NNLO. Eq.~(\ref{decay1}) neglects corrections 
from higher dimension operators. They are incorporated in the numerical 
result below.

The short-distance coefficient is poorly convergent in the 
$\overline{\rm MS}$ factorization scheme implied by the threshold 
expansion. Numerically, at the NRQCD matching scale $\mu_h=m_Q$,
\begin{eqnarray}
\label{cs}
&&\hspace*{-1cm}C(\alpha_s;m_Q)=1-0.849\,\alpha_s(m_Q)
\nonumber\\[0.1cm]
&&\hspace*{-0.5cm}-\,(4.51-0.042 n_f)\,
\alpha_s(m_Q)^2+\ldots,
\end{eqnarray}
where $n_f$ refers to all lighter flavours, approximating them as 
massless. This provided the first hint that NNLO corrections to 
$Q\bar{Q}$ systems are very large \cite{BSS98}. 

The coefficient function is scheme-\-dependent and 
the large NNLO 
correction may be a scheme artefact. For perturbative $Q\bar{Q}$ 
systems one can also compute $|\Psi(0)|^2(\mu)$ in PNRQCD perturbation 
theory (which implies treating the Coulomb interaction 
non-perturbatively). The NNLO $\alpha_s^5$ correction to 
$|\Psi(0)|^2(\mu)$ has 
been obtained analytically in \cite{MY98} and has been 
confirmed by \cite{BSS99}. 
(Note, however, that the result is not presented explicitly in 
\cite{BSS99}. In \cite{PP98} a more complicated representation is 
given. According to \cite{PP98} the two representations agree 
numerically.) The decay width 
for $\Upsilon(1S)\to l^+ l^-$ to 
NNLO in PNRQCD perturbation theory, including the higher dimension 
operators mentioned above, reads 
\begin{eqnarray}
\label{upswidth}
\Gamma(\Upsilon(1S)\to l^+ l^-) &=& \frac{32}{27}\, e_b^2\alpha^2 
m_b \alpha_s(\mu)^3 \Big[1+
\nonumber\\[0.2cm]
&&\hspace*{-2.8cm}
\big(\!-\!1.99 \,l-2.00\big)\,\alpha_s(\mu)+\big(2.64 \,l^2+3.26 \,l
\nonumber\\[0.2cm]
&&\hspace*{-2.8cm}
+\,11.19+
7.43\,\ln(m_b/\mu)\big)\,\alpha_s(\mu)^2\ldots
\Big],
\end{eqnarray}
where $l$ is defined as in (\ref{ups1s}). The logarithm of $m_b/\mu$ 
is partly related to the fact that the natural scale in the short-distance 
coefficient is $m_b$ and not $\mu_1$. However, a consistent treatment 
of all such logarithms requires renormalization group methods and has 
not yet been given. For $\mu_1$ such that $l=0$ (with $m_b=5\,$GeV), 
the factor in brackets is
\begin{equation}
\label{fac1}
F(\mu_1)=1-0.60+1.63.
\end{equation}
As in (\ref{ups1s}) the series is not convergent at all. We have been able 
to eliminate such behaviour as unphysical for the quarkonium mass, but 
we cannot apply the same reasoning here. It therefore seems that 
the leptonic decay width cannot be predicted reliably in perturbation 
theory. 

Despite (\ref{cs}) bottomonium and charmonium decays may well be 
reliably predicted in NRQCD with the NRQCD matrix elements treated as 
non-perturbative parameters. To obtain a definite conclusion one would 
have to compute another quarkonium decay, such as $\eta_b\to 
\gamma\gamma$ or $\Upsilon(1S)\to\,$light hadrons, to NNLO. The ratio of 
decay rates is given by a factorization scheme independent ratio of 
short-distance coefficients. It may be that such ratios are more convergent 
than (\ref{cs}).

A subset of next-to-next-to-next-to-leading order 
contributions to the $\Upsilon(1S)$ 
leptonic decay width is known. 
The ultrasoft correction at order $\alpha_s^6$ contributes 
\begin{eqnarray}
\label{usest3}
\delta F^{us} &=& -6.81\alpha_s^3
\left[\ln\left(\frac{9\mu_{us}}{8 m_b\alpha_s^2}\right)-0.777\right]
\nonumber\\[0.1cm]
&&\hspace*{-1.0cm}\approx\, (-0.15)-(+0.09).
\end{eqnarray}
to $F$ in (\ref{fac1}) \cite{KP99a}. The numerical range is computed with 
the same prescription used for (\ref{usest}). The leading logarithmic 
contribution at order $\alpha_s^6$ is a double logarithm. It arises 
as a product of an ultrasoft or potential logarithm and 
a logarithm related to the anomalous dimension of the current 
$\psi^\dagger\sigma^i\chi$. The correction is \cite{KP99b}:
\begin{eqnarray}
\label{usest4}
\delta F^{ln} &=& -\frac{212}{9\pi}\,
\alpha_s^3 \ln(1/\alpha_s)^2
\nonumber\\[0.1cm]
&& \hspace*{-1.0cm}\approx\,-(0.08-0.30).
\end{eqnarray}
The numerical range is computed with 
the same prescription used for (\ref{usest2}). Again these corrections 
are not small, but their impact is less severe given the already large 
uncertainty of $F(\mu_1)$ in (\ref{fac1}).

In conclusion, higher order corrections almost always turn out to be 
large. The $\Upsilon(1S)$ mass may be useful to determine the $b$ quark 
mass. A reliable prediction of absolute decay widths appears improbable, 
but it remains to be checked whether ratios behave better. This conclusion 
may be frustrating. However, the numerical analysis of the quarkonium mass 
and wave function at the origin provides important insight into the 
structure of corrections to inclusive top and bottom quark pair production 
near threshold.

\section{Top quark pair production near threshold}

Top quark pair production is one of the major 
physics cases for a first linear $e^+ e^-$ collider and has been 
studied extensively in this context \cite{tesla}. The threshold 
behaviour of the cross section can be used to determine 
the top quark mass with great precision -- provided the 
theoretical prediction is accordingly accurate.

Toponium would be the perfect candidate for perturbative applications 
of non-relativistic QCD, but nature has provided another complication. 
The electroweak decay width $\Gamma_t\approx\Gamma(t\to b W)$ increases 
as $m_t^3$. In the standard model $\Gamma_t\approx 1.4\,$GeV, of the 
same order of magnitude as the ultrasoft scale of toponium 
\cite{B86}.

Suppose (for a moment) that the top quark is stable and neglect 
the axial-vector coupling to the $Z$ boson, which is suppressed 
near threshold. Then the 
$t\bar{t}$ production cross section is obtained from the 
correlation function 
\begin{eqnarray}
\label{pi}
\Pi_{\mu\nu}(q^2) &=& (q_\mu q_\nu-q^2 g_{\mu\nu})\,\Pi(q^2) 
\nonumber\\
&&\hspace*{-1.5cm}
= \,i\int d^4 x\,e^{i q\cdot x}\,\langle 0|T(j_\mu(x) j_\nu(0))|0\rangle,
\end{eqnarray}
where $j^\mu(x) = [\bar{t}\gamma^\mu t](x)$ is the top quark vector 
current and $s=q^2$ the centre-of-mass energy squared. Defining 
the usual $R$-ratio $R=\sigma_{t\bar{t}}/\sigma_0$ ($\sigma_0=
4\pi \alpha_{em}^2/(3 s)$, where $\alpha_{em}$ is the electromagnetic 
coupling at the scale $2 m_t$), the relation is 
\begin{equation}
R = \frac{4\pi e_t^2}{s}\,(1+a_Z)\,\mbox{Im}\,\Pi^{ii}(s+i\epsilon),
\label{rr1}
\end{equation}
where $e_t=2/3$ is the top quark electric charge in units of the positron 
charge and $a_Z$ accounts for the vector coupling to the $Z$ boson. (For 
the remainder of this section, I set $a_Z=0$ for simplicity.)
Only the spatial components of the currents contribute up to NNLO. 
In the following $m_t$ refers to the top quark pole mass. 

\begin{figure}[t]
   \vspace{-5.2cm}
   \epsfysize=20cm
   \epsfxsize=15cm
   \centerline{\epsffile{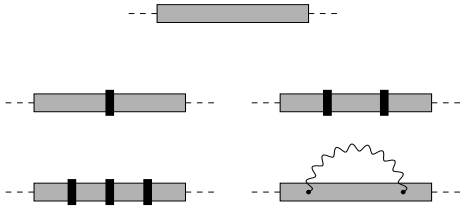}}
   \vspace*{-11.6cm}
\caption[dummy]{\label{fig4}\small PNRQCD perturbative diagrams for 
the heavy quark current correlation function. At leading order the 
current generates a $Q\bar{Q}$ pair which propagates with the Coulomb 
Green function (first line). Black bars denote insertions of potentials. 
The last diagram contains an ultrasoft gluon exchange. Both diagrams 
in the last line are beyond NNLO.}
\end{figure}

At leading order in PNRQCD perturbation theory the heavy quark 
current two-point function is given by the first diagram in 
figure~\ref{fig4}. The $t\bar{t}$ pair is created by the current, 
interacts instantaneously through the (leading order) Coulomb 
potential, and is annihilated by the current. Figure~\ref{fig4} 
is actually misleading. Because $t\bar{t}$ production is 
short-distance compared to the toponium scale $m_t\alpha_s$, 
the pair is created and destroyed at the same space-time point! 
The leading order result is
\begin{equation}
\Pi(s) = \frac{3}{2 m_t^2}\,G_c(0,0;E),
\end{equation}
where $E=\sqrt{s}-2 m_t$. The Green function at the origin is 
ultraviolet divergent. In dimensional regularization, 
with $\overline{\rm MS}$  subtractions, one finds 
\begin{eqnarray}
\label{g0}
G_c(0,0;E) &=& -\frac{m_t^2 \alpha_s}{3\pi}\bigg[
\frac{1}{2\lambda}+\frac{1}{2}\,\ln\frac{-4 m_t E}{\mu^2}
\nonumber\\
&&\hspace*{-1.5cm}-\,\frac{1}{2}+
\gamma_E+\psi(1-\lambda)\bigg],
\end{eqnarray}
where $\lambda=2\alpha_s/(3\sqrt{-E/m_t})$ and $\gamma_E$ is Euler's 
constant. The cross section requires only the 
discontinuity of $\Pi$, which is scheme-independent. The Green 
function contains a continuum at $E>0$ and poles at $E<0$, from 
which the energy and wave function at the origin of toponium resonances 
can be extracted.

How does the top decay width affect this result? For 
top quarks with energy $E\sim m_t v^2\sim\Gamma_t$ and 
momentum $\vec{p}\sim m_t v$, we can approximate the quark 
propagator (in the potential region) 
\begin{eqnarray}
\label{widtht}
\frac{1}{\not\!P_t-m_t-\Sigma(P_t)}&\approx& 
\nonumber\\
&&\hspace*{-2.5cm}\frac{1}{E+i\Gamma_t-
\vec{p}^{\,2}/(2 m_t)} \,\left[1 + {\cal O}(v)\right].
\end{eqnarray}
As expected, the width is a leading order effect, but 
can be taken into account at this order by substituting 
$E\to \bar{E}\equiv E+i\Gamma_t$, where $\Gamma_t$ is the 
gauge-independent on-shell decay width. This 
gives the classic leading order result of \cite{FK87}. Because the 
Green function is evaluated off the real axis, the toponium 
poles are smeared out. For $\Gamma_t\approx 1.4\,$GeV, we 
expect to see a broad remnant of the $1S$ pole, but all higher 
resonances overlap and form a continuum. 

Next-to-leading order corrections to this result have been 
known for some time \cite{SP91}. Other properties of the production 
process, such as top quark momentum distributions and 
asymmetries generated by interference of vector and axial-vec\-tor 
contributions, have been studied in some detail \cite{Kwo91}, 
sometimes with non-perturbative modifications of the heavy quark 
potential that have little justification in the theoretical framework 
described in earlier sections. The recent development concerns the 
calculation of NNLO corrections in a systematic non-relativistic 
approach \cite{HT98,BSS99,NOS99,HT99,PP99}. In PNRQCD Feynman 
diagrams these corrections are given by the two diagrams in the 
second line of figure~\ref{fig4} with potentials up to NNLO. These 
diagrams correspond to integrals of the form
\begin{eqnarray}
&&\hspace*{-0.65cm}
\int\! \frac{d^3\vec{p}}{(2\pi)^3} \frac{d^3\vec{p}^{\,\prime}}{(2\pi)^3} 
\frac{d^3\vec{q}_1}{(2\pi)^3}\frac{d^3\vec{q}_2}{(2\pi)^3} \,\,
\tilde{G}_c(\vec{p},\vec{q}_1;\bar{E})\nonumber\\[0.1cm]
&&\hspace*{0.0cm}
\cdot\,\delta V(\vec{q}_1-\vec{q}_2)
\cdot \,\tilde{G}_c(\vec{q}_2,\vec{p}^{\,\prime};\bar{E})
\end{eqnarray}
and generalizations with more than one insertion of an interaction 
potential $\delta V$. Triple 
insertions of potentials and ultrasoft gluon exchange are 
higher order and neglected. The new calculations 
either employ a combination of numerical and analytical methods 
\cite{HT98,NOS99,HT99} or are fully analytical \cite{BSS99,PP99}. 
The numerical solution of the Schr\"odinger equation contains 
higher order corrections, because an infinite number of insertions of 
the potentials that are kept in the equation is included. 
This could be an advantage if it could be argued that these higher order 
corrections are the dominant ones.

\begin{figure}[t]
   \vspace{-2.3cm}
   \epsfysize=9.5cm
   \epsfxsize=7cm
   \centerline{\epsffile{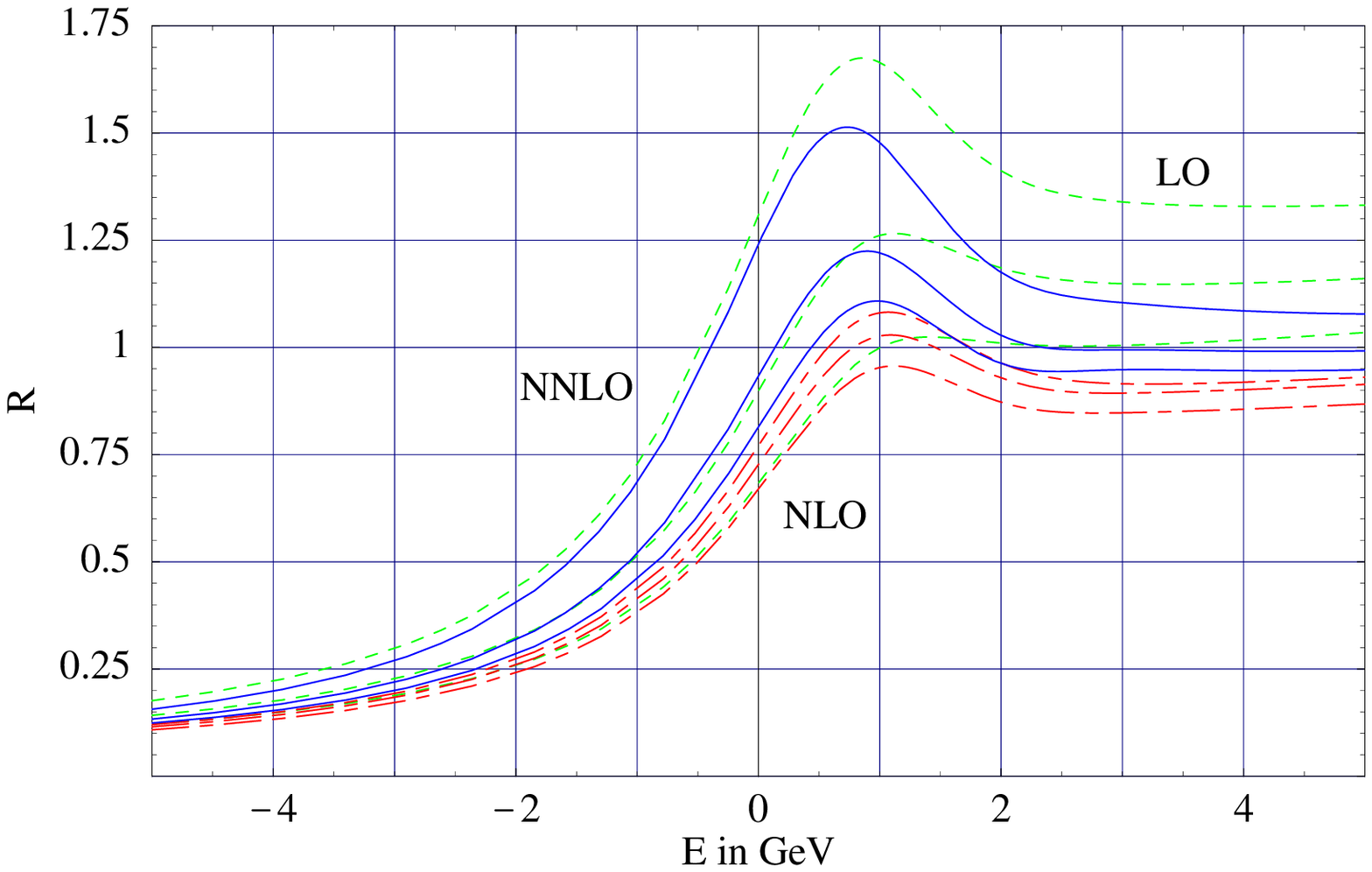}}
   \vspace*{-5.0cm}
   \hspace*{-0.1cm}
   \epsfysize=9.5cm
   \epsfxsize=7cm
   \centerline{\epsffile{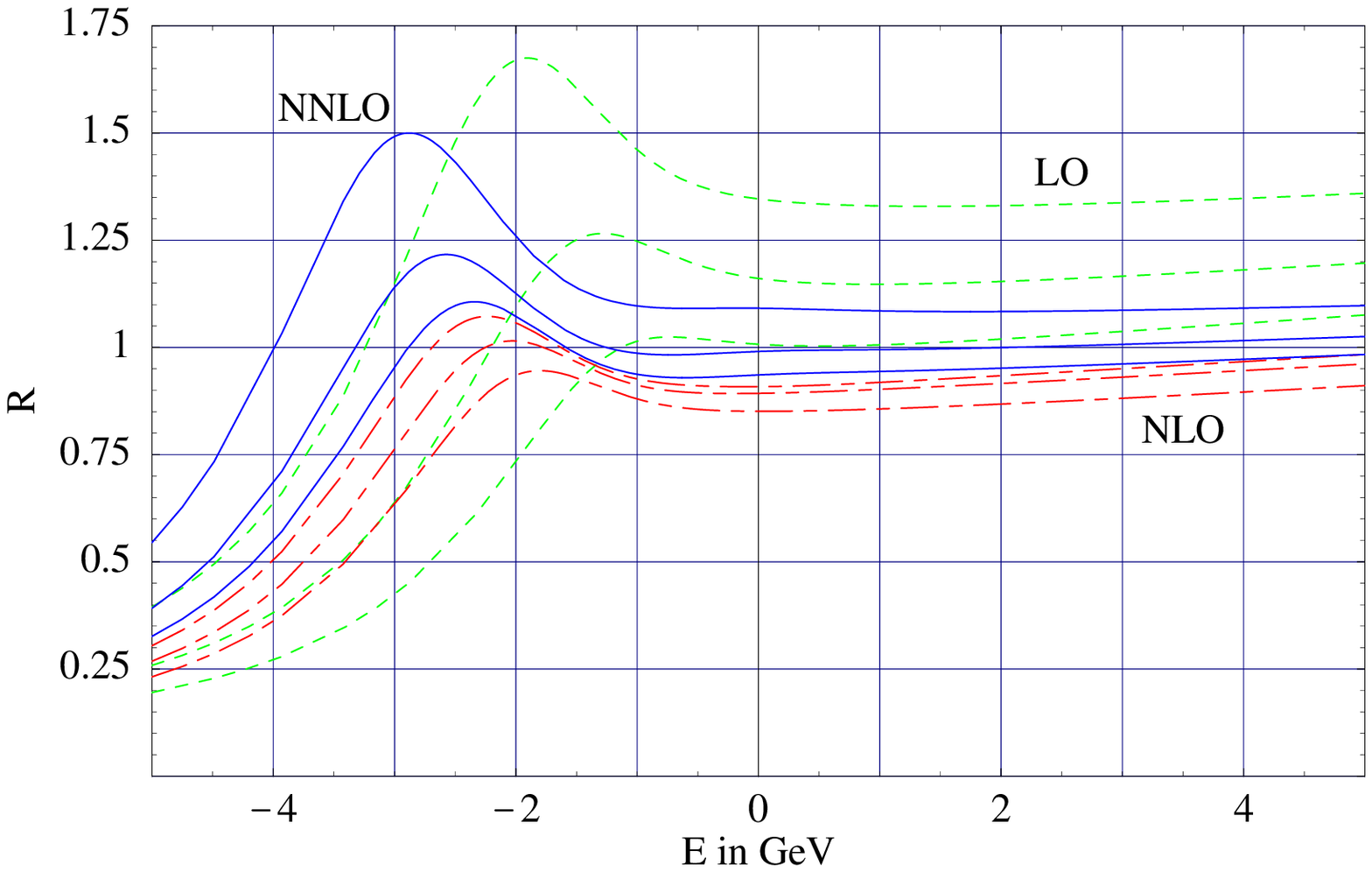}}
   \vspace*{-3.1cm}
\caption[dummy]{\label{fig5}\small 
(a) [upper panel]: The normalized $t\bar{t}$ cross section (virtual 
photon contribution only) in LO (short-dashed), NLO (short-long-dashed) 
and NNLO (solid) as function of 
$E=\sqrt{s}-2 m_{t,\rm PS}(20\,\mbox{GeV})$ (PS scheme, $\mu_f=20\,$GeV).
Input parameters: $m_{t,\rm PS}(20\,\mbox{GeV})=\mu_h=175\,$GeV, 
$\Gamma_t=1.40\,$GeV, 
$\alpha_s(m_Z)=0.118$. The three curves for each case refer to 
$\mu=\left\{15 (\mbox{upper}); 30 (\mbox{central}); 
60 (\mbox{lower})\right\}\,$GeV. (b) [lower panel]: 
As in (a), but in the pole mass scheme. Hence 
$E=\sqrt{s}-2 m_t$. Other parameters as 
above with $m_{t,\rm PS}(20\,\mbox{GeV})\to m_t$. Plot from \cite{BSS99}}
\end{figure}

The first NNLO calculations \cite{HT98} reported large corrections 
to the peak position and normalization of the remnant of the $1S$ 
toponium resonance. The correction to the peak position could be 
explained as an artefact of on-shell renormalization of the top quark 
mass \cite{Ben98}. Subsequent calculations incorporated this observation 
\cite{BSS99,NOS99,HT99} and used one or the other of the 
``alternative'' mass definitions discussed in 
Sect.~\ref{quarkmasses}. The NNLO pair production cross section 
near threshold in the PS scheme is shown 
in the upper panel of figure~\ref{fig5}. For comparison the result 
in the pole mass scheme is shown in the lower panel -- for the very 
last time!  

The generic features of figure~\ref{fig5} are easily 
understood in terms of the results of Sect.~\ref{quarkonium}, since 
the behaviour of successive perturbative approximations in the 
vicinity of the peak reflects essentially the perturbative expansion 
of the toponium $1S$ mass and wave function at the origin. The 
difference in the shift of the peak position in the pole and PS 
scheme is a direct consequence of the improved convergence of the 
perturbative expansion of $M_{1S}$. The stability of the peak position 
in the PS scheme implies that the PS mass (but not the pole mass) 
can be determined accurately from the measurement of the cross 
section. The PS mass determined in this way can also be related 
more reliably to the top quark $\overline{\rm MS}$ mass, which is 
probably the most useful reference parameter. The numerics of 
table~\ref{tab1} adapted to the top quark gives, for given 
$\overline{m}_t=165\,$GeV (and $\alpha_s(\overline{m}_t)=0.1083$):
\begin{eqnarray}
&&\hspace*{-0.65cm}
m_t = \big[165.0+7.58+1.62+0.51
\nonumber\\
&&+\,0.24\,(\mbox{est.})\big]\,
\mbox{GeV}
\label{polerel}\\
&&\hspace*{-0.65cm}
m_{t,\rm PS}(20\,\mbox{GeV}) = \big[165.0+6.66+1.20+
0.28\nonumber\\
&&+\,0.08\,(\mbox{est.})\big]\,
\mbox{GeV},
\label{psrel}
\end{eqnarray}
where the numbers refer to successive terms in the perturbative 
expansion. The difference in convergence is significant on the scale 
of $0.1\,$GeV set by the projected statistical uncertainty on the mass 
measurement. 

However, the perturbative expansion (\ref{upswidth}) 
for the wave function at the 
origin squared remains poorly convergent even for quarks as heavy as top. This 
leads to a large modification of the normalization of the cross section 
near the resonance peak at NNLO and also to a large renormalization 
scale dependence since the NNLO correction is proportional to 
$\alpha_s^5$. The recent calculations \cite{BSS99,NOS99,HT99} agree 
qualitatively on the behaviour of the peak position and normalization. 
When the Schr\"odinger equation is solved numerically, the 
scale dependence of the peak normalization appears to be smaller than in 
figure~\ref{fig5} \cite{HT99}. It is an open question which of the 
two scale dependences provides a realistic estimate of the present 
theoretical uncertainty.

A word of reservation applies to the term ``NNLO''. All current NNLO 
calculations account for the width of the top quark by the 
replacement $E\to E+i\Gamma_t$ or a prescription of similar parametric 
accuracy. Beyond a leading order treatment of (\ref{widtht}), 
counting $\Gamma_t\sim m_t v^2$, the self-energy has to be matched to 
better accuracy. The correction terms relate to the off-shell self-energy 
and carry electroweak gauge-dependence. A complete NNLO result in 
the presence of a width that scales as above therefore includes 
electroweak vertex corrections as well as single resonant backgrounds 
and non-factorizable corrections to the physical $WWb\bar{b}$ 
final state. Although some non-factorizable corrections are known 
near threshold \cite{PS97} and away from threshold \cite{MY94}, 
a systematic treatment of these complications has not been 
attempted yet. Strictly speaking, the concept of the $t\bar{t}$ cross 
section is not defined at NNLO and 
the problem has to be formulated in terms 
of a particular final state such as $WWb\bar{b}$. One may expect 
that single-resonant and non-factorizable corrections are `structureless', 
that is, do not exhibit a pronounced resonance peak. In this case, 
they would add to the already existing normalization uncertainty, but 
would affect little the top quark mass measurement.

Besides the total cross section, top quark momentum distributions 
\cite{NOS99,HT99} and vector-axial-vector interference \cite{PP99} 
have been investigated. These quantities are even more delicate in the 
presence of a finite top quark width, which deserves further 
investigation. Top quark production near threshold in $\gamma\gamma$ 
collisions was considered in \cite{PP98b}. The theoretical accuracy is 
less in this case, because the two-loop short-distance coefficient 
has not yet been calculated.

How large could the corrections to figure~\ref{fig5} be? A 
particularly interesting set of corrections is again related to the 
ultrasoft scale. If we define the ultrasoft scale as the scale 
where the logarithm in (\ref{usest}) vanishes, we obtain 
$\mu_{us}\approx 3\,$GeV for top quarks. 
The scale $\mu_1$ at which $l$ in 
(\ref{ups1s}) vanishes is $32.6\,$GeV. I then repeat 
the estimates (\ref{usest}), (\ref{usest2}), (\ref{usest3}) and 
(\ref{usest4}) for top quarks, using again the results of \cite
{BPSV99b,KP99a,KP99b}, and varying $\mu_{us}$ between 
$2\,$GeV and $5\,$GeV. I obtain  
\begin{eqnarray}
\delta M_{1s}^{us} &\approx& -(140-300)\,\mbox{MeV}
\\
\delta F^{us} &\approx& 0.01-0.05
\end{eqnarray}
for the ultrasoft correction and 
\begin{eqnarray}
\delta M_{1s}^{ln} &\approx& (150-160)\,\mbox{MeV}
\\
\delta F^{ln} &\approx& -0.07
\end{eqnarray}
for the leading logarithmic NNNLO correction. These numbers are relevant 
to $t\bar{t}$ production in the vicinity of the resonance peak. The 
corrections to the peak position are again not small compared to 
the residual uncertainty of about $200\,$MeV in figure~\ref{fig5}, 
but the two corrections are of opposite sign and a conclusive estimate 
requires further NNNLO terms. In addition, one may worry about 
finite width effects in ultrasoft contributions.

To conclude this section, let me make the following remark. It is often 
said that the width of the top quark screens non-perturbative QCD effects 
and makes the threshold cross section calculable in perturbative QCD. 
It is true that the width smears the toponium resonances and converts 
the observed cross section into a smooth excitation curve. However, it 
is not true that the top quark width screens non-perturbative effects 
any more than the existence of a perturbative ultrasoft scale 
$m_t \alpha_s^2\sim \Gamma_t$ already does. Even for stable top quarks 
the production cross section near threshold 
(averaged over an interval several times 
$\Lambda_{\rm QCD}$) is perturbatively calculable, as would be 
the toponium resonances and their decays. Interestingly, perturbative 
resummation with power counting $v\sim \alpha_s$ seems to make sense 
even as $v\to 0$ at fixed $\alpha_s$. In particular, the cross section 
directly at threshold seems to be infrared safe in perturbation theory. 
As $v\to 0$ the arguments of the 
coupling constants $\alpha_s(m_Q v)$ and 
$\alpha_s(m_Q v^2)$ freeze at values of order $m_Q\alpha_s$ and 
$m_Q\alpha_s^2$, respectively, and do not tend to zero. This has been 
checked explicitly by investigating the logarithms up to order 
$\alpha_s^3$ (NNLO).

\section{Sum rules, the $b$ quark mass}

As a final application we return to the $b$ quark mass. There are 
legitimate doubts concerning the reliability of (\ref{mass1}) given 
that $\Upsilon(1S)$ is hardly a truly perturbative onium. One can 
by-pass this problem by considering averages over the bottom pair 
production cross section rather than exclusive resonances. This 
leads us to consider sum rules \cite{NSVZ77}
\begin{eqnarray}
\label{eq1}
M_n/(10\,\mbox{GeV})^{2 n}&\equiv& \frac{12\pi^2}{n!}\,
\frac{d^n}{d(q^2)^n}\,\Pi(q^2)_{\big| q^2=0} 
\nonumber\\
&&\hspace*{-1.5cm}= 
\int\limits_0^\infty \!
\frac{ds}{s^{n+1}}\,R_{b\bar{b}}(s)
\end{eqnarray}
which equate an experimental average of the cross section to the
perturbatively computable derivatives (``moments'') 
of the bottom vector current correlation function. 

The parameters of the lowest $\Upsilon(nS)$ resonances are well 
measured and one chooses $n$ large enough that the experimental 
uncertainty on the continuum cross section is small compared to the 
theoretical uncertainty. This occurs for $n\geq 6$. For such moments 
the ordinary perturbative expansion of the moments breaks down and has 
to be replaced by non-relativistic resummation and PNRQCD perturbation 
theory. Leading order and next-to-leading order analyses of the 
sum rule with non-relativistic resummation have been performed 
in \cite{VZ87} and \cite{Vol95}, respectively. After integration over 
$s$ the expansion of $R_{b\bar{b}}(s)$ in $\alpha_s/v$ turns into 
an expansion in $\alpha_s\sqrt{n}$. Non-relativistic resummation 
sums these terms to all orders. For moments the scale $m_b v$ turns 
into $2 m_b/\sqrt{n}$; the ultrasoft scale $m_b v^2$ is $m_b/n$. 
The requirement that the ultrasoft scale is perturbative translates 
into $n\leq 10$. Larger moments are often used in the literature. This 
introduces a systematic uncertainty which is difficult to quantify 
at NNLO, since at this order ultrasoft corrections are not 
included.

Several NNLO calculations have been completed recently 
\cite{PP98,Hoa98,MY98,Hoa99,BS99}, the calculation being almost 
identical to that of top pair production near threshold. The 
later publications \cite{MY98,Hoa99,BS99} abandoned the idea 
of determining the pole mass and usually extract the bottom 
$\overline{\rm MS}$ mass $\overline{m}_b$. This is done by extracting 
the PS or 1S or kinetic mass from the sum rule, which is then 
converted into $\overline{m}_b$. Although the different groups compute 
the same quantity, there are differences in the implementation 
of the resummation which are formally beyond NNLO. These concern 
(a) whether the short-distance coefficient (\ref{cs}) is kept as 
an overall factor or multiplied out to NNLO; (b) whether the integral 
over $s$ in (\ref{eq1}) is done exactly or in a non-relativistic 
approximation; (c) whether the energy denominator of the full 
Green function is expanded up to NNLO or whether the exact NNLO 
energy levels are kept. These differences in implementation 
can shift the value of $m_b$ extracted from the sum rule by up to 
$100\,$MeV.

\begin{figure}[h]
   \vspace{-1.9cm}
   \epsfysize=9cm
   \epsfxsize=6.8cm
   \centerline{\epsffile{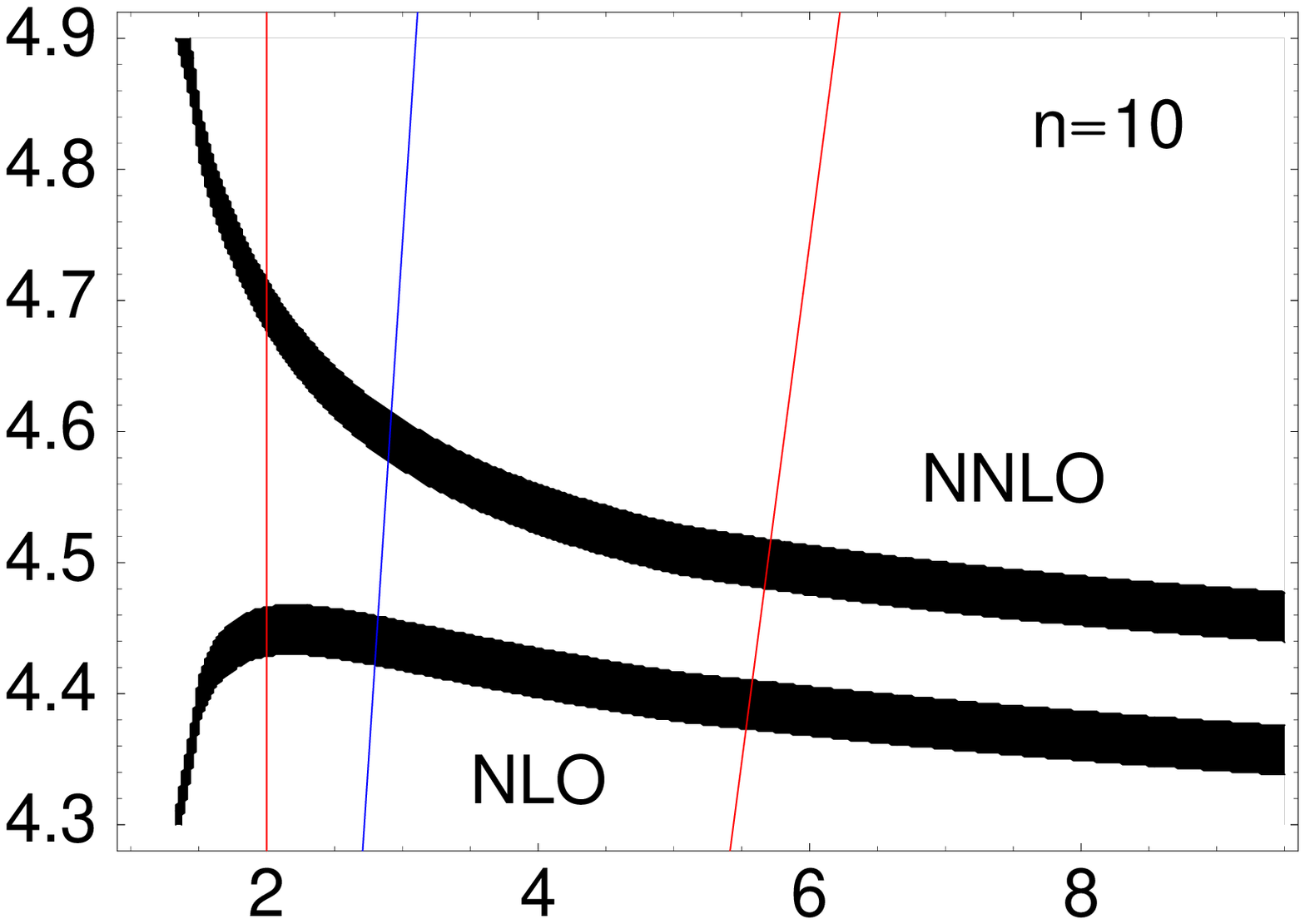}}
   \vspace{-3.9cm}
   \epsfysize=9cm
   \epsfxsize=6.8cm
   \centerline{\epsffile{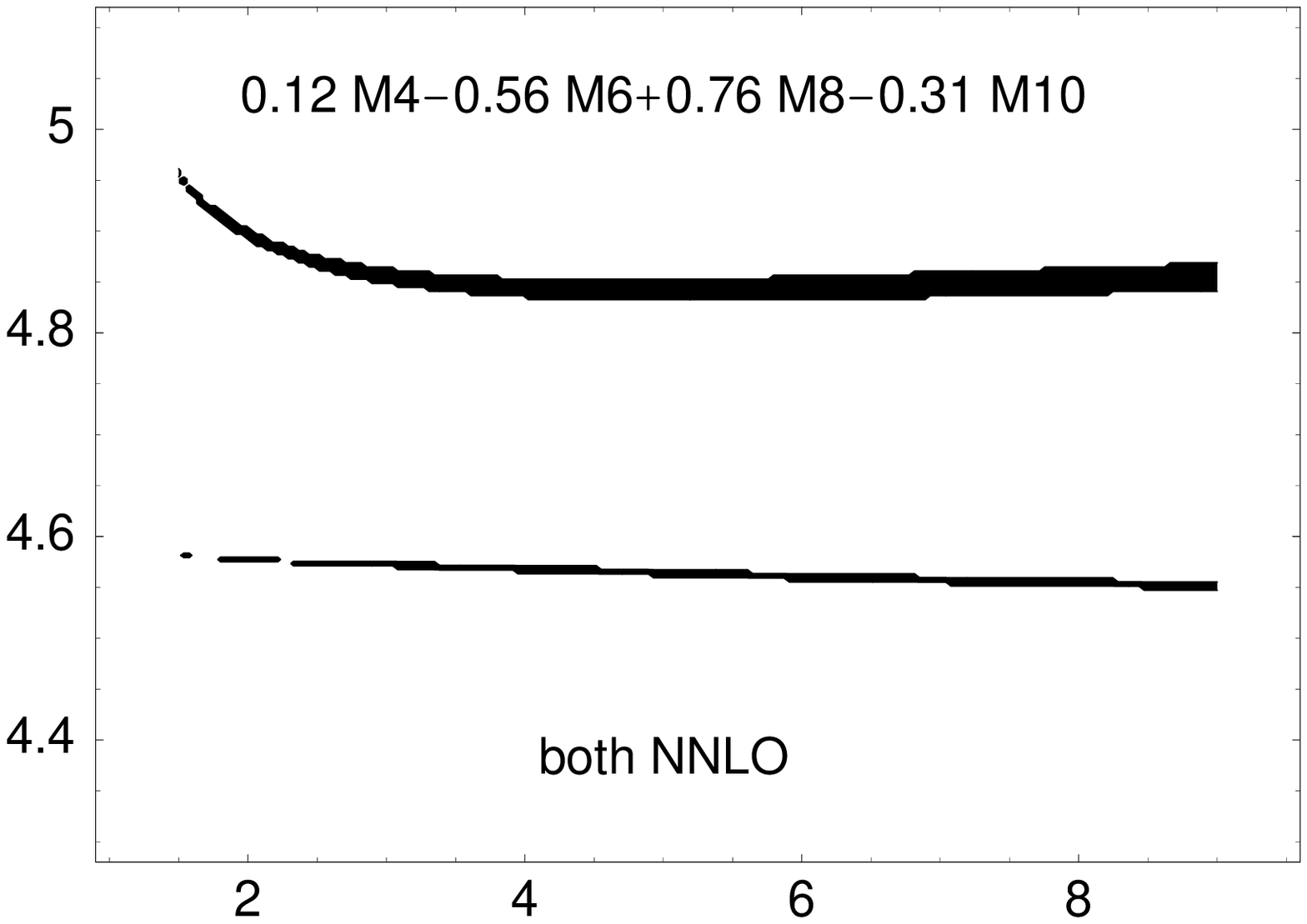}}
   \vspace*{-2.3cm}
\caption[dummy]{\label{fig6}\small 
(a) [upper panel]: The value of $m_{b,\rm PS}(2\,\mbox{GeV})$ 
obtained from the 10th moment as a function of the renormalization scale 
in NLO and NNLO and for $\alpha_s(m_Z)=0.118$. 
The dark region specifies the variation due to the 
experimental error on the moment. The middle line marks the scale 
$\mu_n=2 m_b/\sqrt{n}$, 
the two outer lines determine the scale variation from which the 
theoretical error is computed. Plot from \cite{BS99}. 
(b) [lower panel]: 
The value of $m_{b,\rm PS}(2\,\mbox{GeV})$ 
obtained from a linear combination of moments 
as a function of the renormalization scale 
in NNLO. $M_n$ is defined with the normalization of (\ref{eq1}).}
\end{figure}

Further differences arise in the choice of moments, renormalization scale, 
at which the sum rule is evaluated, and the analysis strategy.

Refs.~\cite{MY98,BS99} perform essentially a single-moment analysis, 
based on the fact that the theoretical error is highly correlated 
between different moments. It is then checked that varying the moment 
gives a negligible difference. One then finds a significant 
renormalization scale uncertainty, which can be traced back to the 
badly behaved expansion (\ref{fac1}). A typical result is shown 
in the upper graph of figure~\ref{fig6}. The results 
\begin{eqnarray}
&&\hspace*{-0.8cm}
m_{b,\rm PS}(2\,\mbox{GeV}) = (4.60\pm 0.11)\,\mbox{GeV}
\quad\!\!\cite{BS99}
\\
\label{kin}
&&\hspace*{-0.8cm}
m_{b,\rm kin}(1\,\mbox{GeV}) = (4.56\pm 0.06)\,\mbox{GeV}
\quad\!\!\!\cite{MY98}
\end{eqnarray}
differ by about $80\,$MeV, when related to each other according to 
table~\ref{tab1}, but are consistent with each other within 
implementational differences. The larger error on the first result 
follows from a larger renormalization scale variation.

\TABULAR{|c|c|c|c|}{
\hline 
&&&\\[-0.4cm]
Refs. & $\overline{m}_b(\overline{m}_b)$ & $m_b$ & Remarks\\[0.1cm] 
\hline
&&&\\[-0.4cm]
\cite{MY98}  & $4.20\pm 0.10$ & $--$ & Sum rules 
(via $m_{b,\rm kin}(1\,\rm GeV)$) \\
&&&\\[-0.4cm]
\cite{Hoa99,hoa99b} & $4.19\pm 0.06$ & $--$ & Sum rules (via $m_{b,1S}$)\\
&&&\\[-0.4cm]
\cite{BS99,BSS98II} & $4.26\pm 0.10$ & $4.97\pm 0.17$ & Sum rules 
(via $m_{b,\rm PS}(2\,\rm GeV)$)  \\
&&&\\[-0.4cm]
\cite{PP98} & $4.21\pm 0.11$  & $4.80\pm 0.06$ & Sum rules (via 
$m_b$, 2-loop) \\
&&&\\[-0.4cm]
\hline
&&&\\[-0.4cm]
\cite{PY98} & $4.44\pm 0.04$ & $5.04\pm 0.09$ & 
$\Upsilon(1S)$\,mass (via $m_b$, 2-loop) \\
&&&\\[-0.4cm]
\cite{BS99} & $4.24\pm 0.09$ & $--$ & 
$\Upsilon(1S)\,$mass (via $m_{b,\rm PS}(2\,\rm GeV)$) \\
&&&\\[-0.4cm]
\cite{hoa99b} & $4.21\pm 0.07$ & $--$ & 
$\Upsilon(1S)\,$mass (direct) \\[-0.4cm]
&&&\\
\hline}
{Bottom quark $\overline{\rm MS}$ and pole mass values (in GeV) obtained 
from NNLO sum rule or $\Upsilon(1S)$ mass calculations. 
\label{tab3}}

The analysis of \cite{Hoa99} is different, because it uses linear 
combinations of moments. The linear combination chosen in \cite{Hoa99} 
is less sensitive to a variation of the renormalization scale, while 
retaining sensitivity to $m_b$. This is illustrated in the lower panel of 
figure~\ref{fig6} in the PS scheme. In this way, \cite{Hoa99} obtains
\begin{equation}
\label{1s}
m_{b,1S} = (4.71\pm 0.03)\,\mbox{GeV},
\end{equation}
which is close to (\ref{kin}) after conversion. This  
result depends crucially on combining moments 
at identical renormalization scales rather than their ``natural'' scale 
$2 m_b/\sqrt{n}$, and on discarding possible multiple solutions 
(as the upper one in figure~\ref{fig6}b). In my opinion the 
error quoted in (\ref{1s}) should be understood as an 
error that follows under the specific assumptions 
of the analysis strategy. It can hardly be considered as a 
realistic estimate of the total theoretical error, given the 
differences that can 
arise in different implementations of the theoretical moments and given 
the size of ultrasoft effects discussed in Sect.~\ref{masses}. 

The results quoted above can be converted into the bottom 
$\overline{\rm MS}$ mass using table~\ref{tab1}. A summary of NNLO 
results from sum rules and, for comparison, the $\Upsilon(1S)$ mass, 
is compiled in table~\ref{tab3}, where I quote the number 
given by the authors. This number may differ from the one of 
table~\ref{tab1}, because not always is a four-loop related to 
$\overline{m}_b$ used as appropriate to a NNLO sum rule calculation 
(cf. the discussion after (\ref{mass1})). This difference is 
small when mass definitions with convergent relations to 
$\overline{m}_b$ are used, see table~\ref{tab1}, but affects 
$\overline{m}_b$ when computed from the pole mass via a 2-loop 
relation. For this reason the results for $\overline{m}_b$ 
from \cite{PY98,PP98} in table~\ref{tab3} should in fact 
be decreased by about $200\,$MeV. This makes \cite{PY98} consistent 
with the other $\overline{m}_b$ determinations, but puts the 
result of \cite{PP98} off by $200\,$MeV. Comparison of pole mass 
results shows that this discrepancy is already present in the 
pole mass value before conversion to $\overline{m}_b$. The small 
pole mass value of \cite{PP98} is a consequence of the fact that 
the sum rule is evaluated at high renormalization scales. 
Fig.~\ref{fig5}a shows that this leads to reduction of $m_b$. 

My ``best''estimate for the $b$ quark $\overline{\rm MS}$ mass is a 
(potentially biased) combination of the results of 
\cite{MY98,Hoa99,BS99}. It is remarkable that the sum rule 
result is consistent with the $\Upsilon(1S)$ results despite the 
fact that the systematics of non-perturbative effects and 
scale dependence is different. Nevertheless, the results of 
table~\ref{tab3} are not independent and may be affected by common, 
unidentified theoretical uncertainties. This being said, my 
preferred value is
\begin{equation}
\overline{m}_b(\overline{m}_b) = (4.23\pm 0.08)\,\mbox{GeV}.
\end{equation}
This is in beautiful agreement with 
$\overline{m}_b(\overline{m}_b) = 4.26 \pm 0.07\,\mbox{GeV}$ 
\cite{GGRM99} obtained from lattice HQET. This uses the $B$ meson mass, 
a lattice calculation of the (properly defined) binding energy of the 
$B$ meson in the unquenched, two-flavour approximation to heavy quark 
effective theory, and a two-loop perturbative matching to the 
$\overline{\rm MS}$ scheme. To our knowledge, this is the only 
other NNLO determination of the $\overline{\rm MS}$ mass besides 
the sum rule calculations mentioned above (which, in fact, are N$^4$LO as 
far as $\overline{m}_b$ is concerned). In my opinion using an 
error smaller than $80\,$ MeV on $\overline{m}_b(\overline{m}_b)$ 
or any of the ``alternative'' masses cannot be justified at 
present. Using smaller errors in $B$ physics observables may have dangerous 
consequences for consistency checks of the CKM model of CP 
violation.

\section{Concluding remarks}

The theory of perturbative onium systems has undergone an exciting 
transition from potential models with arbitrary cut-offs and 
poorly understood accuracy to a systematic effective theory 
description. This 
transition can be compared in significance with the development 
of heavy quark effective theory for  $B$ and $D$ mesons. Unfortunately, 
nature has not been kind to us, leaving us with systems which are 
barely perturbative (bottomonium) or extremely short-lived 
(toponium). 

Along with this development perturbative expansion tools have been 
invented and the most basic quantities are now computed to 
next-to-next-to-leading order. These calculations sharpened 
our understanding of heavy quark mass re\-nor\-malization, but large 
corrections remain in most cases. They confront us with 
the challenge of a complete 
next-to-next-to-next-to-leading order calculation. 
With so many tools at hand, progress is surely expected.

\acknowledgments
I would like to thank the organizers of HF8 for making this a focused 
and enjoyable meeting. Thanks to A.~Signer and V.A.~Smirnov for 
collaboration on subjects related to this review and to A.H.~Hoang 
and A.~Pineda for discussions. This work was supported in part by the 
EU Fourth Framework Programme `Training and Mobility
of Researchers', Network `Quantum Chromodynamics and the Deep Structure 
of Elementary Particles', contract FMRX-CT98-0194 (DG 12 - MIHT).

\end{document}